\documentclass[usenatbib]{mn2e}

\input{psfig.sty}
\usepackage{graphicx}

\def\beq{\begin{equation}}
\def\enq{\end{equation}}

\title{Long-term Hard X-ray Monitoring of 2S 0114+65 with INTEGRAL/IBIS}

\author[Wei Wang]{Wei Wang\thanks{E-mail: wangwei@bao.ac.cn} \\
National Astronomical Observatories, Chinese Academy of Sciences,
Beijing 100012, China}

\begin{document}
\maketitle

\begin{abstract}

We present the results of the long-term hard X-ray monitoring of
the high mass X-ray binary 2S 0114+65 with INTEGRAL/IBIS from 2003
to 2008. 2S 0114+65 is a variable hard X-ray source with X-ray
luminosities of $10^{35} - 4\times 10^{36}$ erg s$^{-1}$ from 20
-- 100 keV due to accretion rate changes in different orbital
phases. In several observations when 2S 0114+65 was bright, we
found a pulse period evolution of $\sim 2.67$ hour to 2.63 hour
from 2003 -- 2008, with a spin-up rate of the neutron star $\sim
1.09\times 10^{-6}$ s s$^{-1}$. Compared with the previous
reported spin-up rate, the spin-up rate of the neutron star in 2S
0114+65 is accelerating. The spectral properties of 2S 0114+65 in
the band of 18 -- 100 keV which changed with the orbital phases,
generally could be described with a power-law model with a high
energy exponential cutoff. The variation of the power-law photon
index over orbital phase anticorrelates with hard X-ray flux, and
the variation of $E_{\rm cut}$ has a positive correlation with the
hard X-ray flux, implying that the harder spectrum at the maximum
of the light curve. The variations of spectral properties over
orbital phase suggested 2S 0114+65 as a highly obscured binary
system. In some observational revolutions, hard X-ray tails above
70 keV are detected. We study the characteristics of the hard
X-ray tails combining JEM-X and IBIS data in the energy range of 3
-- 100 keV. The 3 -- 100 keV spectra of 2S 0114+65 are generally
fitted by an absorbed power-law model with high energy cutoff. We
discover that the hard X-ray tails are only detected when column
density is very low. Thus, high column density leads to
disappearance of the hard X-ray tails in this wind-fed neutron
star accretion binary. Our results would help to understand the
origin, evolution and properties of this peculiar class of
super-slow pulsation neutron stars in high mass X-ray binaries.

\end{abstract}


\section{INTRODUCTION}

The high mass X-ray binary 2S 0114+65 is an unusual system, which
shows properties consistent with both Be and supergiant X-ray
binaries, with significant temporal and spectral variability over
a wide range of time scales (Crampton et al. 1985). A type B1a
supergiant optical counterpart (LS I+65010) at a distance of $\sim
7.2$ kpc was identified (Reig et al. 1996). 2S 0114+65 has an
orbital period of $\sim 11.59$ day (Crampton et al. 1985; Corbet
et al. 1999; Wen et al. 2006; Grundstrom et al. 2007). A
superorbital modulation at $\sim 30.7$ day was also reported from
the analysis of 8.5 years of RXTE/ASM data (Farrell et al. 2006).
The low X-ray luminosity of $\sim 10^{36}$ erg s$^{-1}$ (in the
3-- 20 keV band, Hall et al. 2000) implies that spherical
accretion takes place via the stellar wind of the donor star in 2S
0114+65 (Li \& van den Huevel 1999). It should be noted that the
He II 4686 \AA line (a common signature of the presence of an
accretion disc) has been observed in some occasions but was very
weak (Aab et al. 1983, Crampton et al. 1985), possibly linked to
the presence of a small transient accretion disc.

The X-ray source 2S 0114+65 consists of an wind-fed accretion
neutron star with its pulse period first measured at $\sim 2.78$
hr by Finley et al. (1992). Hall et al. (2000) derived a period of
$\sim 2.73$ hr, and found a spin-up rate of $\sim 6.2\times
10^{-7}$ s s$^{-1}$ for the neutron star in 2S 0114+65 over $\sim
11$ yr. Using the INTEGRAL/IBIS data, Bonning \& Falanga (2005)
obtained a spin period of $\sim 2.67$ hr, and a spin-up rate of
$\sim 8.9\times 10^{-7}$ s s$^{-1}$ from 1996 to 2004. Farrell et
al. (2008) analyzed pointed observation data of RXTE around early
2006 to derive a spin period of $\sim 2.65$ hr. From 2004 to 2006,
the spin-up rate of 2S 0114+65 was $\sim 1.6\times 10^{-6}$ s
s$^{-1}$ if we compare the results by INTEGRAL (Bonning \& Falanga
2005) and RXTE (Farrell et al. 2008). Then the spin-up rate seems
to be accelerating in recent years though the pulsed period was
derived from different instruments. The reported long spin period
in 2S 0114+65 makes it one of the slowest X-ray pulsars. Many
theoretical models has been suggested to explain this long period
pulsation (Hall et al. 2000; Li \& van den Heuvel 1999; and
references therein). Li \& van den Heuvel (1999) suggested the
magnetar origin for the neutron star in 2S 0114+65 to explain the
slow pulsation. At present, the magnetic field of the neutron star
in 2S 0114+65 has not been determined. The possible electron
cyclotron resonant absorption features at $\sim 22$ keV and 44 keV
was reported by Bonning \& Falanga (2005), implying a magnetic
field of $\sim 2.5\times 10^{12}$ G.

2S 0114+65 as an accretion neutron star system has the typical
X-ray spectrum (i.e., 1 -- 50 keV) fitted with an absorbed
power-law model with the high energy exponential cut-off (Farrell
et al. 2008; Masetti et al. 2006; Hall et al. 2000). Recently, den
Hartog et al. (2006) found the high energy tail above 70 keV in 2S
0114+65 using the INTEGRAL/IBIS data. The origin of the hard X-ray
tail in 2S 0114+65 is unclear. In addition, we need to study the
hard X-ray spectra up to 100 keV to confirm the existence of the
hard X-ray tails in 2S 0114+65.

In this paper, we present our results of long-term high energy
monitoring of 2S 0114+65 up to $\sim 100$ keV with INTEGRAL/IBIS
observations from 2003 to 2008. Our scientific studies mainly aim
to discover the different hard X-ray properties of 2S 0114+65 in
different accretion states, search for the hard X-ray tails and
possible electron cyclotron absorption feature in the spectra and
derive the pulse period of 2S 0114+65 in different time intervals
to probe the spin-up rate of the neutron star from 2003 to 2008.
Observations of IBIS and science data analysis were described in
\S 2. In \S 3., we show the timing analysis and results of spin
period evolution. The spectral properties in different accretion
states/orbital phases and searching for hard X-ray tails and
possible cyclotron absorption features were presented in \S 4.
Summary and discussions of the results were given in the last
section.

\begin{figure}
\centering
\includegraphics[angle=0,width=9cm]{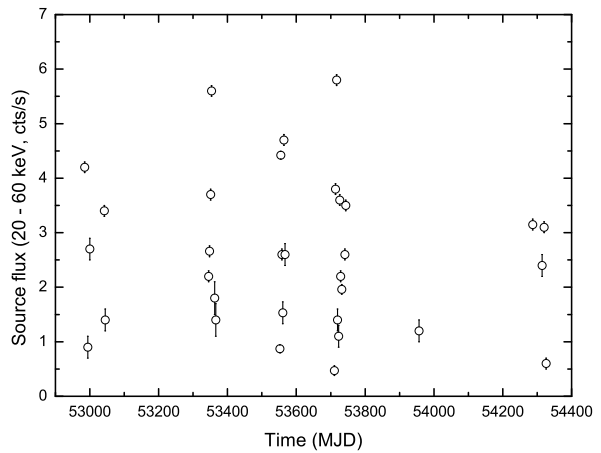}
\caption{The light curve of the high massive X-ray binary 2S
0114+65 binning at $\sim 3$ days in the energy range of 20-- 60
keV observed by IBIS from 2003 Dec to 2008 May.  }
\end{figure}

\begin{figure}
\centering
\includegraphics[angle=0,width=8.5cm]{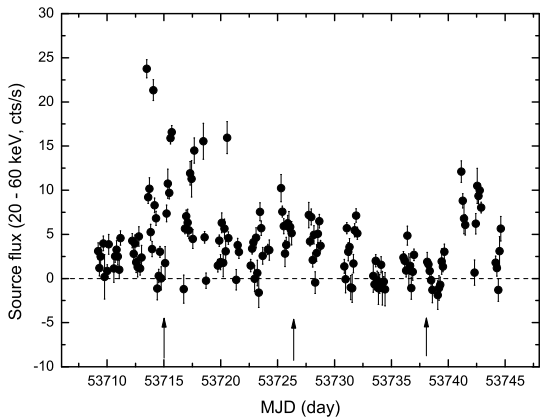}
\caption{Hard X-ray light curve of 2S 0114+65 observed by
INTEGRAL/IBIS from the revolutions 384 -- 395 with 0.15 day
binning. For a comparison, we also show the maximum flux from 1.5
-- 12 keV based on the RXTE/ASM data which is indicated with the
arrows.  }
\end{figure}

\section{Observations and data analysis}

\begin{table*}

\caption{INTEGRAL/IBIS observations of the field around 2S
0114+65. The time intervals of observations in the revolution
number and the corresponding dates, the corrected on-source
exposure time, mean off-axis angle on the source are listed. And
mean count rate and the detection significance level value in the
energy range of 20 -- 60 keV were also shown. }

\begin{center}
\scriptsize
\begin{tabular}{l c c c c l}

\hline \hline Rev. Num. & Date  & On-source time (ks) & off-axis angle & Mean count rate s $^{-1}$ & Detection level \\
\hline 142 & 2003 Dec 12--14 & 99 & 7.9$^\circ$ & 4.1$\pm 0.1$ & 33$\sigma$ \\
143 & 2003 Dec 15--17 & 94 & 7.7$^\circ$& $<0.6$ & $<5\sigma$  \\
144  & 2003 Dec 18 -- 20 & 90 & 7.8$^\circ$ & $<0.7$ & $<5\sigma$ \\
145  & 2003 Dec 21 --23 & 93 &9.8$^\circ$ & $0.9\pm 0.2$ & 5$\sigma$  \\
146 & 2003 Dec 24 --26 & 58 &6.5$^\circ$ & $<0.7$ & $<5\sigma$ \\
147 & 2003 Dec 27 --38 & 92 & 9.7$^\circ$ & $2.9\pm 0.2$ & 15$\sigma$\\
148 & 2003 Dec 30 -- 2004 Jan 01  & 91 &7.9$^\circ$&  $<0.6$ & $<5\sigma$ \\
161 &  2004 Feb 7 -- 9 & 75 & 9.7$^\circ$ &$3.4\pm 0.1$ & 33$\sigma$ \\
162 & 2004 Feb 10 -- 12 & 78 &9.8$^\circ$ & $1.4\pm 0.2$ & 7$\sigma$ \\
238 & 2004 Sep 24 -- 26 & 198 & 1.8$^\circ$ & $<0.4$ & $<5\sigma$ \\
262 & 2004 Dec 6 -- 7 & 63 & 2.1$^\circ$ & 2.2$\pm 0.1$ & 19.6$\sigma$ \\
263 & 2004 Dec 8 -- 10 & 76 & 1.9$^\circ$ & 2.66$\pm 0.09$ & 27$\sigma$ \\
264 & 2004 Dec 11 -- 13 & 97 & 8.5$^\circ$ &3.7$\pm 0.1$ & 35$\sigma$ \\
265 & 2004 Dec 14 -- 16 & 103 &  8.7$^\circ$ &5.6$\pm 0.1$ & 48$\sigma$\\
266 & 2004 Dec 17 -- 19 & 86 & 9.1$^\circ$ & $<0.7$ & $<5\sigma$ \\
268 & 2004 Dec 20 -- 22 & 45 & 8.9$^\circ$ & $1.8\pm 0.3$ & 6$\sigma$ \\
269 & 2004 Dec 23 -- 25 & 81 & 8.8$^\circ$ &$1.4\pm 0.3$ & 5$\sigma$ \\
331 & 2005 June 29 -- July 2 & 190 & 3.1$^\circ$ &$0.87\pm 0.07$ & $12.9\sigma$ \\
332 & 2005 Jul 2 -- 5 & 173 & 3.4$^\circ$ &$4.42\pm 0.07$ & $61.4\sigma$ \\
333 & 2005 Jul 5 -- 7 & 206 & 3.3$^\circ$ &$2.60\pm 0.06$ & $40.5\sigma$ \\
334 & 2005 Jul 8 -- 10 & 60 & 4.3$^\circ$ &$1.53\pm 0.13$ & $11.3\sigma$ \\
335 & 2005 Jul 11 -- 13 & 170 & 4.6$^\circ$ &$4.69\pm 0.08$ & $59.1\sigma$ \\
336 & 2005 Jul 14 -- 16 & 104 & 5.3$^\circ$ &$2.56\pm 0.17$ & $15.3\sigma$ \\
384 & 2005 Dec 5 -- 7 & 99  &  8.1$^\circ$ &$0.47\pm 0.09$ & 5$\sigma$ \\
385 & 2005 Dec 8 -- 10 & 98 & 9.3$^\circ$ &$3.9\pm 0.1$ & 37$\sigma$\\
386 & 2005 Dec 11 -- 13 & 101 & 9.0$^\circ$ &5.8$\pm 0.1$ & 55$\sigma$ \\
387 & 2005 Dec 14 -- 16 & 90 & 8.7$^\circ$ &$1.4\pm 0.2$ & 7$\sigma$ \\
388 & 2005 Dec 17 -- 19 & 93 & 8.3$^\circ$ &$1.1\pm 0.2$ & 5$\sigma$ \\
389 & 2005 Dec 20 -- 22 & 108 & 7.9$^\circ$ &$3.6\pm 0.1$ & 35$\sigma$ \\
390 & 2005 Dec 23 -- 25 & 117 & 7.7$^\circ$ &2.2$\pm 0.1$ & 23$\sigma$ \\
391 & 2005 Dec 26 -- 28 & 93 & 8.4$^\circ$ &1.96$\pm 0.09$ & 19$\sigma$ \\
392 & 2005 Dec 29 -- 31 & 105 & 8.6$^\circ$ &$<0.3$ & $<5\sigma$ \\
393 & 2006 Jan 1 -- 3 & 88 & 8.5$^\circ$ &$<0.4$ & $<5\sigma$\\
394 & 2006 Jan 4 -- 6 & 102 & 8.0$^\circ$ &$2.6\pm 0.1$ & 25$\sigma$\\
395  & 2006 Jan 7 -- 9 & 98 & 7.8$^\circ$ &$3.5\pm 0.1$ & 36$\sigma$\\
454  & 2006 Jul 4 -- 6 & 208 & 1.3$^\circ$ &$< 0.4$ & $<5\sigma$ \\
466 & 2006 Aug 7 -- 9 & 78 & 8.9$^\circ$ &1.2$\pm 0.2$ & 7$\sigma$ \\
528 & 2007 Feb 8 -- 10 & 196 & 1.5$^\circ$ &$< 0.4$ & $<5\sigma$ \\
664 & 2008 Mar 21 -- 23 & 95 & 1.6$^\circ$ &3.15$\pm 0.10$ & 30$\sigma$  \\
667 & 2008 Mar 30 -- Apr 01 &46 & 1.4$^\circ$  & $< 0.6$ & $<5\sigma$ \\
668 & 2008 Apr 2 -- 4 & 44 & 1.4$^\circ$ &$< 0.6$ & $<5\sigma$ \\
669 & 2008 Apr 5 -- 7 & 44 & 1.1$^\circ$ &$< 0.5$ & $<5\sigma$ \\
670 & 2008 Apr 8 -- 10 & 47 & 0.9$^\circ$ &$< 0.5$ & $<5\sigma$ \\
673 & 2008 Apr 17 -- 19 & 50 & 0.9$^\circ$ &$2.37\pm 0.14$ & $16.5\sigma$ \\
675 & 2008 Apr 23 -- 25 & 109 & 0.7$^\circ$ &$3.09\pm 0.09$ & $39.2\sigma$ \\
677 & 2008 Apr 29 -- May 1 & 96 & 0.8$^\circ$ &$0.60\pm 0.09$ & $6.3\sigma$ \\
 \hline
\end{tabular}
\end{center}

\end{table*}

The INTErnational Gamma-Ray Astrophysics Laboratory (INTEGRAL,
Winkler et al. 2003) is ESA's currently operational space-based
hard X-ray/soft gamma-ray telescope. There are two main
instruments aboard INTEGRAL, the imager IBIS (Ubertini et al.
2003) and the spectrometer SPI (Vedrenne et al. 2003),
supplemented by two X-ray monitors JEM-X (Lund et al. 2003) and an
optical monitor OMC (Mas-Hesse et al. 2003). All four instruments
are co-aligned, allowing simultaneous observations in a wide
energy range. SPI has a lower sensitivity than IBIS for weak
source below 100 keV, and detection of weak point sources is not
so significant for spectral studies. Here we use the data
collected with the low-energy array called IBIS-ISGRI (INTEGRAL
Soft Gamma-Ray Imager) which consists of a pixellated 128$\times
128$ CdTe solid-state detector that views the sky through a coded
aperture mask (Lebrun et al. 2003). IBIS/ISGRI has a 12' (FWHM)
angular resolution and arcmin source location accuracy in the
energy band of 15 -- 200 keV. JEM-X collects the lower energy
photons from 3 -- 35 keV. Since JEM-X has a small field of view
(like requiring off-axis angle $<5^\circ$) and a relatively low
sensitivity because of small detector area (Lund et al. 2003), the
source 2S 0114+65 was only significantly detected in a few time
intervals. We use the JEM-X data combined with IBIS observations
in some special time intervals to constrain the hard X-ray tails
from the broadband spectra.

The high mass X-ray binary 2S 0114+65 was observed frequently
during the INTEGRAL surveys of the Cassiopeia region. In this
work, we use the available archival data for the IBIS observations
where 2S 0114+65 was within $\sim 10$ degrees of the pointing
direction. In Table 1, we summarize the INTEGRAL/IBIS observations
in our analysis, including the revolution numbers, corrected
exposure time and off-axis angle on the source for each
revolution. The archival data used in our work are available from
the INTEGRAL Science Data Center (ISDC).

\begin{figure*}
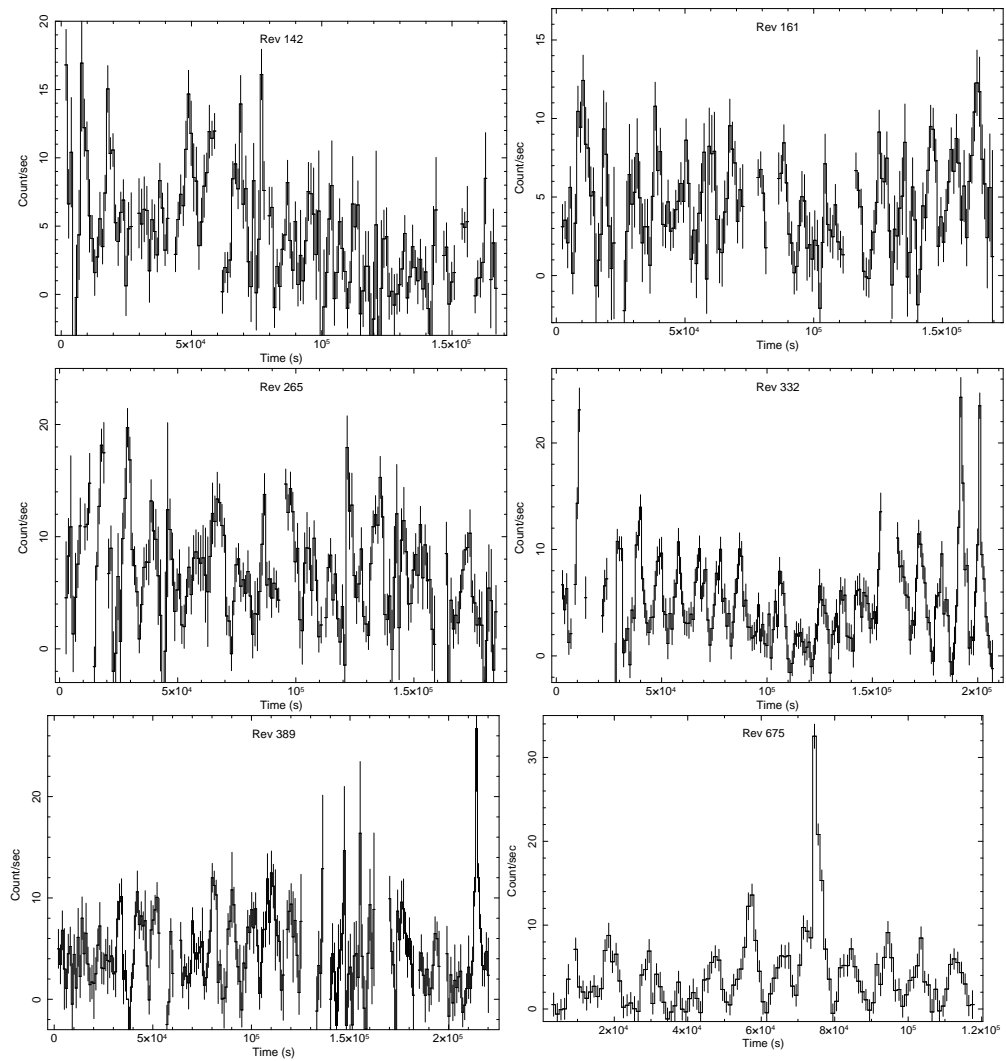

\centering
\includegraphics[angle=-90,width=6.5cm]{2s0114_lc_142.ps}
\includegraphics[angle=-90,width=6.5cm]{2s0114_lc_161.ps}
\includegraphics[angle=-90,width=6.5cm]{2s0114_lc_265.ps}
\includegraphics[angle=-90,width=6.5cm]{2s0114_lc_332.ps}
\includegraphics[angle=-90,width=6.5cm]{2s0114_lc_389.ps}
\includegraphics[angle=-90,width=6.7cm]{2s0114_lc_675.ps}
\caption{IBIS background subtracted hard X-ray light curves of 2S
0114+65 in the energy band of 20 -- 60 keV for six different time
intervals: Rev 142 (2003 Dec), 161 (2004 Feb), 265 (2004 Dec), 332
(2005 July), 389 (2005 Dec) and 675 (2008 April).}
\end{figure*}

\begin{figure*}
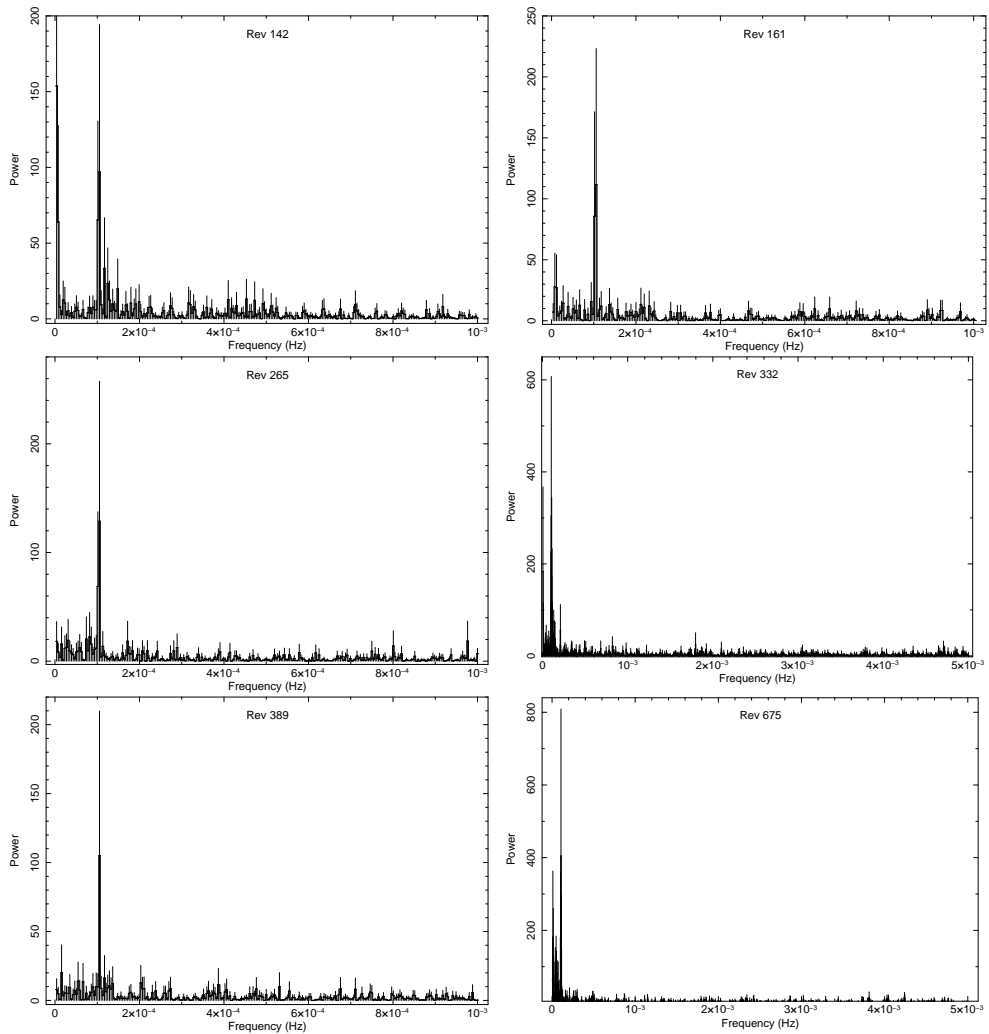

\centering
\includegraphics[angle=-90,width=6.5cm]{2s0114_pw_142.ps}
\includegraphics[angle=-90,width=6.5cm]{2s0114_pw_161.ps}
\includegraphics[angle=-90,width=6.5cm]{2s0114_pw_265.ps}
\includegraphics[angle=-90,width=6.5cm]{2s0114_pw_332.ps}
\includegraphics[angle=-90,width=6.5cm]{2s0114_pw_389.ps}
\includegraphics[angle=-90,width=6.5cm]{2s0114_pw_675.ps}
\caption{Power spectra of hard X-ray light curves of 2S 0114+65
for six different time intervals in Fig. 3: Rev 142 (2003 Dec),
161 (2004 Feb), 265 (2004 Dec), 332 (2005 July), 389 (2005 Dec)
and 675 (2008 April).}
\end{figure*}

\begin{figure*}
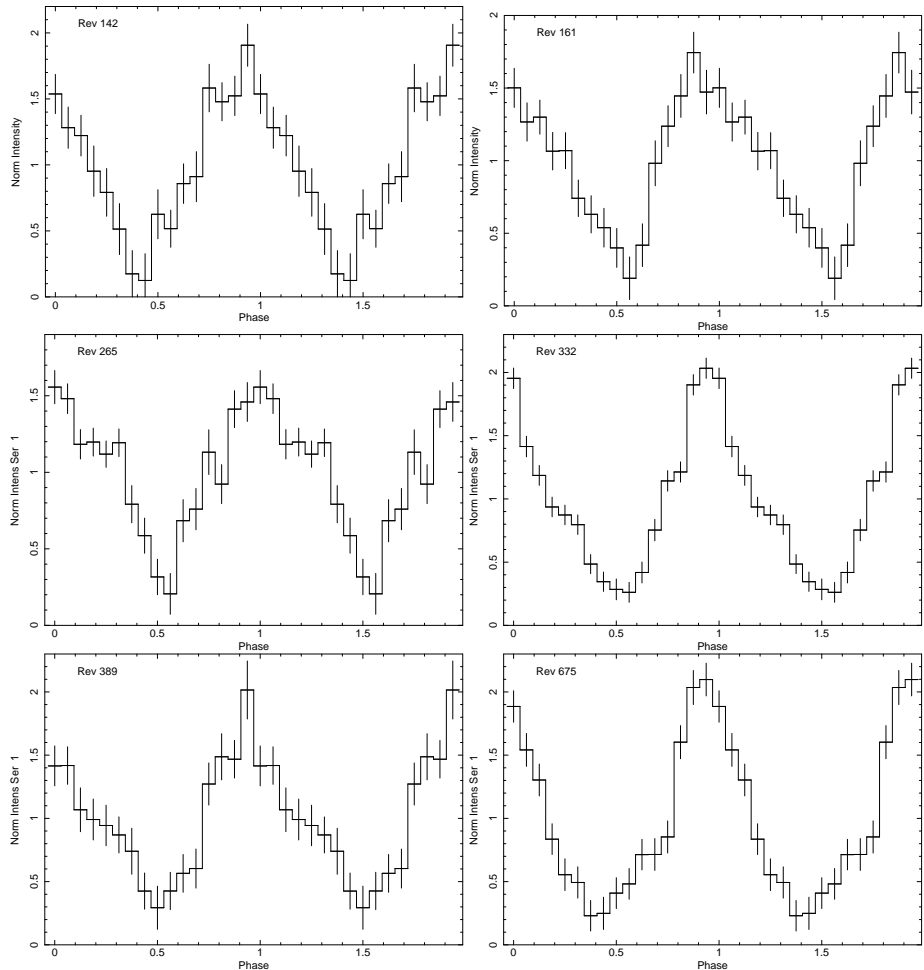

\centering
\includegraphics[angle=-90,width=6cm]{2s0114_pul_142.ps}
\includegraphics[angle=-90,width=6cm]{2s0114_pul_161.ps}
\includegraphics[angle=-90,width=6cm]{2s0114_pul_265.ps}
\includegraphics[angle=-90,width=6cm]{2s0114_pul_332.ps}
\includegraphics[angle=-90,width=6cm]{2s0114_pul_389.ps}
\includegraphics[angle=-90,width=6cm]{2s0114_pul_675.ps}
\caption{The IBIS/ISGRI background subtracted light curves (20 --
60 keV) of 2S 0114+65 folded at a pulsation period for six
different time intervals: Rev 142 ($P=9612$ s), 161 ($P=9600$ s),
265 ($P=9570$ s), 332 ($P=9555$ s), 389 ($P=9520$ s), 675
($P=9475$ s). The pulse profiles are repeated once for clarity. }
\end{figure*}

The analysis was done with the standard INTEGRAL off-line
scientific analysis (OSA, Goldwurn et al. 2003) software, ver.
7.0. Individual pointings processed with OSA 7.0 were mosaicked to
create sky images according to the methods and processes described
in Bird et al (2007). We used the 20 -- 60 keV band for source
detection and to quote source fluxes for each revolution (see
Table 1). In most revolutions, 2S 0114+65 was detected
significantly with IBIS, while for some orbit phases, the source
was not detected (the detection significance level was lower than
5$\sigma$), then only the upper limits of the count rates were
given in Table 1.

In Figure 1, the light curve of 2S 0114+65 from 2003 December to
2006 August in the energy range of 20 -- 60 keV detected by IBIS
was displayed. 2S 0114+65 is a variable source whose mean hard
X-ray flux in each revolution detected by IBIS varied from $\sim
0.4$ cts/s to $\sim 6$ cts/s in the range of 20 -- 60 keV. This
X-ray variability could be due to different accretion states or
accretion in the different orbit phases if the orbit is eccentric.
In this paper we classify the light curve into two accretion
states: quiescent states with average flux below $\sim 1$ cts/s or
non-detection in one revolution observations; and active states.
For these different accretion states, 2S 0114+65 would have
different spectral properties which will be studied in \S 4.

\section{Timing analysis and spin period evolution of 2S 0114+65}

The high massive X-ray binary 2S 0114+65 has an orbital period of
$\sim 11.59$ days according to the regular soft X-ray observations
by RXTE/ASM (Corbet et al. 1999; Farrell et al. 2008). Hard X-ray
observations by IBIS/ISGRI have shown that 2S 0114+65 is a
variable source. Its variability may also be related to the
orbital modulation. IBIS had continuous observations of 2S 0114+65
from 2005 Dec 5 -- 2006 Jan 9. This observational time scale
covered about 3 orbit periods (Figure 2). The hard X-ray light
curves from 20 -- 60 keV of 2S 0114+65 show no obvious orbital
modulation. For a comparison, we also show the epochs of the
maximum flux from 1.5 -- 12 keV derived by RXTE/ASM in the same
time interval from MJD 53700 -- 53750, as indicated with the
arrows.

Our IBIS observations of the X-ray pulsar 2S 0114+65 covered near
5 years, so it is the good way for us to derive the spin period
evolution of 2S 0114+65 from 2003 to 2008 using the present data.

\subsection{Spin period evolution}

To search for the pulsation period and the spin evolution of 2S
0114+65, we use the observational data in which 2S 0114+65 has a
higher detection significant level (e.g., $> 30\sigma$, see Table
1). Then we select the databases for six time intervals covering
nearly 5 years: 2003 Dec 12 -- 14 (Rev 142); 2004 Feb 7 -- 9 (Rev
161); 2004 Dec 14 -- 16 (Rev 265); 2005 Jul 2 -- 5 (Rev 332); 2005
Dec 20 -- 22 (Rev 389) and 2008 Apr 23 -- 25 (Rev 675). The
minimum time bins in analysis are taken as $\sim 200$ s for each
dataset. The hard X-ray light curves in the energy band of 20 --
60 keV obtained by IBIS/ISGRI in six time intervals are presented
in Figure 3 and the barycentric corrections were also made. The
X-ray light curves in Fig. 3 have been re-binned to 1000 s for
clarity. We apply FFT on each of the six light curves to search
for the modulation period signals. The power spectra for each
light curve are displayed in Figure 4. Significant modulation
period signals around $\sim 9500 - 9600$ s are found for all light
curves.

According to the position of peak signals in the power spectra
(Figure 4), we use the pulse-folding technique to derive the pulse
period values of 2S 0114+65 for the six time intervals: $P=9612\pm
20$ s for Rev 142; $P=9600\pm 20$ s for Rev 161; $P=9570\pm 20$ s
for Rev 265; $P=9555\pm 15$ s for Rev 332; $P=9520\pm 20$ s for
Rev 389; and $P=9475\pm 25$ s for Rev 675. Figure 5 shows the
light curves of 2S 0114+65 folded with the derived pulse period in
six time intervals. The pulse light curves show the single main
peak feature. The pulse fraction defined as the ratio of the pulse
maximum minus the minimum to the maximum can also be estimated for
each pulse light curve. We found the pulse fractions of $\sim
(95\pm 5)\%$ for Rev 142, $\sim (89\pm 8)\%$ for Rev 161, $\sim
(88\pm 7)\%$ for Rev 265, $\sim (86\pm 4)\%$ for Rev 332, $\sim
(85\pm 9)\%$ for Rev 389, $\sim (91\pm 6)\%$ for Rev 675.

The obtained pulse period values of 2S 0114+65 suggested that the
X-ray pulsar in 2S 0114+65 is still spinning up. Figure 6
displayed the spin history of 2S 0114+65 from different
measurements. In particular, Finley et al. (1992) derived a spin
period of $\sim 2.78$ hr, and nine years later, Hall et al. (2000)
found a pulse period of $\sim 2.73$ hr. More recently, Bonning \&
Flanga (2005) obtained a pulse period of $\sim 2.67$ hr using IBIS
observations, and Farrell et al. (2008) got a pulse period of
$\sim 2.65$ hr using RXTE data. From their measurements in
different time intervals, we obtain a history of spin-up rates for
the X-ray pulsars in 2S 0114+65: $\sim 6.2\times 10^{-7}$ s
s$^{-1}$ from 1986 -- 1996; $\sim 8.9\times 10^{-7}$ s s$^{-1}$
from 1996 to 2004; $\sim 1.6\times 10^{-6}$ s s$^{-1}$ from 2004
to 2006. Figure 6 also shows six data points using our IBIS
observations from 2003 Dec to 2008 May. From our IBIS measurements
in six time intervals, we can find a spin-up rate of $\sim
(1.09\pm 0.13)\times 10^{-6}$ s s$^{-1}$ from 2003 to 2008.
Therefore, the spin period of the neutron star in 2S 0114+65 was
decreasing, and the spin-up rate was really accelerating in the
past thirty years.

\begin{figure}
\centering
\includegraphics[angle=0,width=9cm]{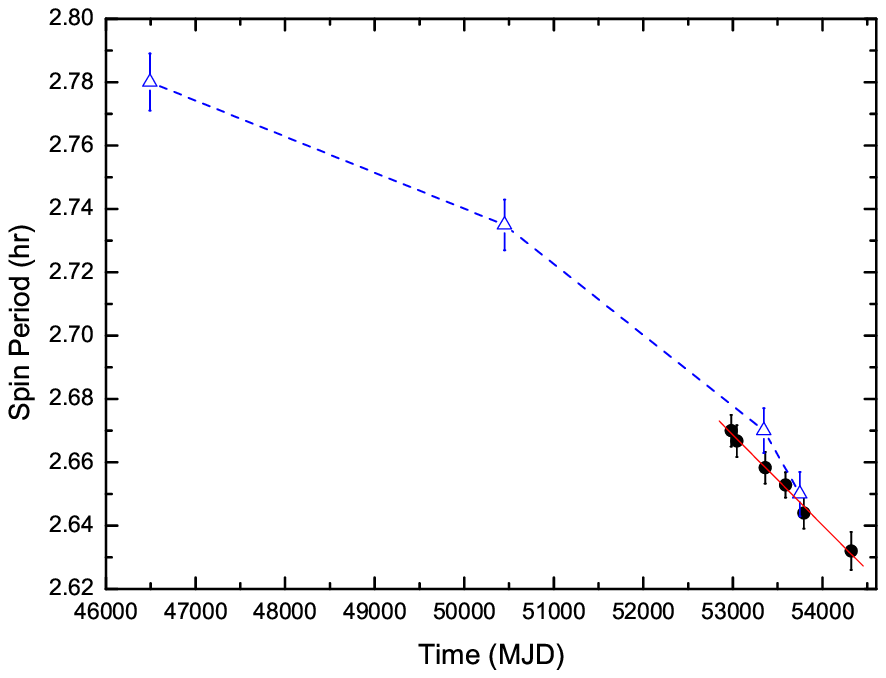}
\caption{Spin evolution history of 2S 0114+65 from different
observations. The filled circles are from this work. The open
triangles are from Finley et al. (1992), Hall et al. (2000),
Bonning \& Flanga (2005) and Farrell et al. (2008). The dashed
lines show the spin-up rates in different time intervals. The
solid line presents the best fitting rate of $\sim (1.09\pm
0.13)\times 10^{-6}$ s s$^{-1}$ from 2003 -- 2008 only using our
IBIS data. See details in the text. }
\end{figure}

\section{Spectral properties in hard X-rays}

Hard X-ray spectral properties (above 15 keV) of 2S 0114+65 have
not been studied in details before. From our analysis of the five
years of hard X-ray monitoring data by INTEGRAL/IBIS, we study the
properties of 2S 0114+65 up to 200 keV. 2S 0114+65 is a highly
variable X-ray source due to accretion variability and different
orbital phases. We then study the hard X-ray spectral properties
of 2S 0114+65 in different accretion states and orbital phases. In
particular, we will try to search for the hard X-ray tails and
possible cyclotron absorption line features around 22 keV and 44
keV in the hard X-ray spectra. This will help us to understand the
formation and evolution of this X-ray source system.

From Table 1 and Figure 1, the X-ray binary 2S 0114+65 has
variable hard X-ray fluxes during the hard X-ray monitoring with
IBIS from 2003 -- 2008. We would show the spectral properties of
2S 0114+65 in different accretion states as defined in \S 2,
namely active states and quiescent states. The spectral
extractions for the observations of each revolution were carried
out. The spectral analysis software package used is XSPEC 12.4.0x.

Spectral properties of 2S 0114+65 could be variable in different
accretion rates, then we first use a simple model like a single
power-law model to fit all the spectra, which show the changes of
the spectra properties for comparison. In Table 2, we show the
spectral properties of 2S 0114+65 in both active states and
quiescent states. In the quiescent cases, we study the spectral
properties in four revolutions (Rev. 145, 331, 384, 677) when the
IBIS rate is below $\sim 1$ cts/s in Table 1. In active states, we
carry out spectral analysis on the source in the revolution when
2S 0114+65 was detected with the significance levels higher than
$\sim 30 \sigma$ in Table 1. Since we also want to check the
possible cyclotron absorption features around 22 keV and 44 keV
reported by Bonning \& Falanga (2005) who used the IBIS data in
Rev. 262 and 263. Then in the active case, we show the spectral
properties of 2S 0114+65 for 15 revolutions in Table 2. To
estimate the X-ray luminosity, we have assumed a source distance
of 7.2 kpc for 2S 0114+65 throughout the paper (Reig et al. 1996).

\begin{table*}

\caption{Spectral properties of 2S 0114+65 in different accretion
states and datasets (for observations in each revolution). All
spectra are fitted with a single power law model for the
comparison. The hard X-ray fluxes and luminosities in the range of
20 -- 100 keV for the active states and 20 -- 50 keV for quiescent
states in model fits were also given.}

\begin{center}
\scriptsize
\begin{tabular}{l c c c l}

\hline \hline Rev. Num. & $\Gamma$ &  Flux ($10^{-10}$ erg cm$^{-2}$ s$^{-1}$) & $L_x$ (erg s$^{-1}$) & reduced $\chi^2$ \\
\hline
{\bf The active states} &  &  &   &  \\
\hline 142 & $2.5\pm 0.1$ & $3.0\pm 0.3$ & 2.1$\times 10^{36}$ & $1.261 (28 d.o.f.)$ \\
161  & $2.8\pm 0.1$ &  $2.3\pm 0.4$ & 1.5$\times 10^{36}$ & $1.252 (25 d.o.f.)$ \\
262  & $2.6\pm 0.2$   &   $1.6\pm 0.3$ & 1.0$\times 10^{36}$ & $0.845 (26 d.o.f.)$ \\
263  & $2.7\pm 0.1$   &   $1.9\pm 0.3$ & 1.2$\times 10^{35}$ & $1.177 (26 d.o.f.)$ \\
264 & $2.7\pm 0.1$  &  $2.5\pm 0.4$ & 1.7$\times 10^{36}$ & $0.796 (24 d.o.f.)$ \\
265  & $2.6\pm 0.1$  &   $3.9\pm 0.3$ & 2.7$\times 10^{36}$ & $1.352 (27 d.o.f.)$ \\
332 & $2.7\pm 0.1$  &   $3.0\pm 0.2$ & 2.1$\times 10^{36}$ & $1.630 (28 d.o.f.)$ \\
333 & $2.7\pm 0.1$  &   $1.8\pm 0.2$ & 1.1$\times 10^{36}$ & $1.951 (28 d.o.f.)$ \\
335 & $2.6\pm 0.1$  &   $3.3\pm 0.3$ & 2.3$\times 10^{36}$ & $1.562 (28 d.o.f.)$ \\
385  & $2.7\pm 0.1$ &  $2.6\pm 0.3$ & 1.7$\times 10^{36}$ & $0.973 (27 d.o.f.)$ \\
386  & $2.6\pm 0.1$ &  $4.1\pm 0.3$ & 2.8$\times 10^{36}$  & $1.490 (29 d.o.f.)$ \\
389  & $2.8\pm 0.1$  &  $2.4\pm 0.3$ & 1.5$\times 10^{36}$ & $1.052 (29 d.o.f)$ \\
395  & $2.6\pm 0.1$  &  $2.4\pm 0.3$ & 1.5$\times 10^{36}$ & $0.986 (29 d.o.f)$ \\
664 & $2.8\pm 0.1$  &   $2.4\pm 0.3$ & 1.5$\times 10^{36}$ & $0.956 (28 d.o.f.)$ \\
675 & $2.7\pm 0.1$  &   $2.6\pm 0.3$ & 1.7$\times 10^{36}$ & $1.356 (28 d.o.f.)$ \\
\hline
{\bf The quiescent states} &   &  &   &  \\
\hline 145  & 2.7$\pm 0.6$ & 0.47$\pm 0.11$ & 2.9$\times 10^{35}$
& $0.689 (5 d.o.f)$ \\
331  & $3.1\pm 0.3$  & $0.51\pm 0.09$ & 3.1$\times 10^{35}$ &$1.52 (4 d.o.f.)$ \\
384  & $2.6\pm 0.7$  & $0.19\pm 0.07$ & 1.2$\times 10^{35}$ & $1.21 (3 d.o.f.)$ \\
677  & $3.2\pm 0.6$ & $0.29\pm 0.09$ & 1.7$\times 10^{35}$ &
$0.953 (4 d.o.f.)$
\\ \hline
\end{tabular}
\end{center}

\end{table*}

\subsection{Spectral properties in different states}

\subsubsection{active states}

During the 5 years of hard X-ray monitoring, the high mass X-ray
binary 2S 0114+65 may undergo active states in most of the time
(see Fig. 1). We will derive the hard X-ray spectral information
for 2S 0114+65 with the data of each revolution in active states,
so that the variations of the spectral properties with time can be
presented. In this subsection, we show the spectral properties of
2S 0114+65 in active states when the source was detected by IBIS
with the significance levels $>30 \sigma$, which includes 13
revolutions (see Table 1). In addition, two revolutions 262 and
263 (detection significance levels are $20\sigma$ and $27\sigma$
respectively) are also included to check the possible absorption
features around 44 keV reported by Bonning \& Falanga (2005).

We fit the spectra with the same model -- a single power-law
model. In Table 2, the detailed fitted spectral parameters for
each observational revolution are presented together for the
comparison. In Figure 7, we display the IBIS-ISGRI spectra of 2S
0114+65 in the active states from nine observations (as examples):
Rev 161; Rev 262; Rev 263; Rev 264; Rev 265; Rev 385; Rev 386; Rev
389 and Rev 395. The spectra in the other revolutions will been
shown in the following sections, so they are not repeated here.

The spectrum of the source 2S 0114+65 has been studied up to 100
keV. But 2S 0114+65 is very weak above 100 keV, and is only
marginally detected in some cases (like very bright, or possible
existence of high energy tails). Then the data points in the
spectra of 2S 0114+65 generally show only upper limits above 100
keV. We present only a few data points above 100 keV to show the
high energy behavior (no detection or possible tails) of 2S
0114+65 (see examples in Fig. 7). In Table 2, we show the hard
X-ray fluxes and derived X-ray luminosity in the range of 20 --
100 keV. In active states, 2S 0114+65 has a hard X-ray luminosity
in the range of $(1-3)\times 10^{36}$ erg s$^{-1}$ from 20 -- 100
keV.

The single power-law model gives the photon index range of 2.5 --
2.8 implying moderate spectral variations in active states. The
given reduced $\chi^2$ in the power-law fittings is generally
around 1, but in some revolutions it is significantly higher than
1 (see Table 2). So the simple power-law model would be unsuitable
for the spectra in these revolutions. In the {\bf Appendix}, we
present an extended table to display spectral parameters for all
revolutions shown in Table 2 fitted with the different models like
a thermal bremsstrahlung model, a power-law model with a high
energy exponential cut-off (hereafter {\em pow*highe}) or the
bremsstrahlung model plus a power-law model (for the possible hard
X-ray tail case). This extended table could help the readers to
justify which model may be more acceptable for each spectrum.

\begin{figure*}
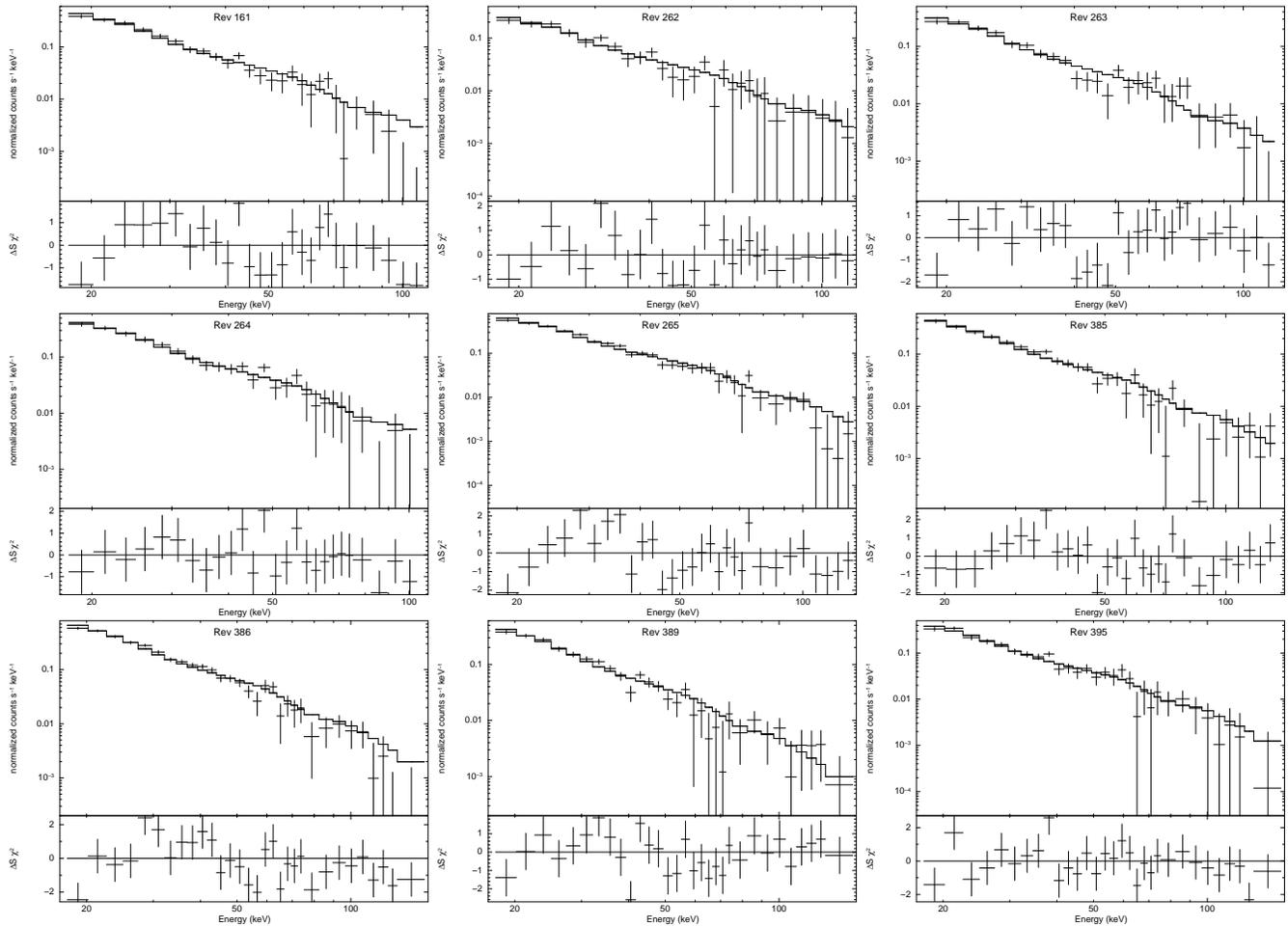

\centering
\includegraphics[angle=-90,width=5.8cm]{2s0114_spe_161.ps}
\includegraphics[angle=-90,width=5.8cm]{2s0114_spe_262.ps}
\includegraphics[angle=-90,width=5.8cm]{2s0114_spe_263.ps}
\includegraphics[angle=-90,width=5.8cm]{2s0114_spe_264.ps}
\includegraphics[angle=-90,width=5.8cm]{2s0114_spe_265.ps}
\includegraphics[angle=-90,width=5.8cm]{2s0114_spe_385.ps}
\includegraphics[angle=-90,width=5.8cm]{2s0114_spe_386.ps}
\includegraphics[angle=-90,width=5.8cm]{2s0114_spe_389.ps}
\includegraphics[angle=-90,width=5.8cm]{2s0114_spe_395.ps}
\caption{The hard X-ray spectra of 2S 0114+65 obtained by
INTEGRAL/IBIS during the observational revolutions in the active
states: 161, 262, 263, 264, 265, 385, 386, 389 and 395. All
spectra are fitted with a single power-law model for comparison.
The spectral properties of 2S 0114+65 evolved in the different
times from our observations. See details of the fitting parameters
in Table 2 and the text. }
\end{figure*}

In active states, 2S 0114+65 was detected in hard X-rays by IBIS
with a high significance level of above $30 \sigma$, which may
help us to search for the possible cyclotron absorption lines
around 22 keV and 44 keV reported by Bonning \& Falanga (2005).
Bonning \& Falanga (2005) found possible absorption feature around
44 keV used the IBIS data during observations in Revs. 262 and
263. We also derived the spectrum of 2S 0114+65 in Revs. 262 and
263 but with the different energy band binning (see Fig. 7). From
residuals after continuum fitting, the absorption feature around
44 keV is not obvious in Rev 262, but might still exist in Rev
263. In the other seven spectra in active states presented in Fig.
7, the possible absorption feature around 44 keV was only hinted
in Rev 265. In addition, possible features were also found around
54 keV in Rev 385 and around 40 keV in Rev 389. So we cannot
confirm the possible cyclotron absorption features around 22 keV
and 44 keV. These possible features in residuals may be induced by
the flux fluctuations derived by the IBIS detectors, possibly due
to effects of detector response and calibration uncertainties.
However, we do not exclude the possible existence of the cyclotron
absorption line around 22 keV in 2S 0114+65, because production of
electron cyclotron absorption lines may depend on the accretion
geometry and the relativistic cyclotron absorption line energy
would have non-harmonic line spacing (Araya \& Harding 1999) and
fundamental line energy also changes in different observations
(depending on luminosity and other relevant effects, see Klochkov
et al. 2008; Araya \& Harding 2000).

\subsubsection{quiescent states}

\begin{figure*}
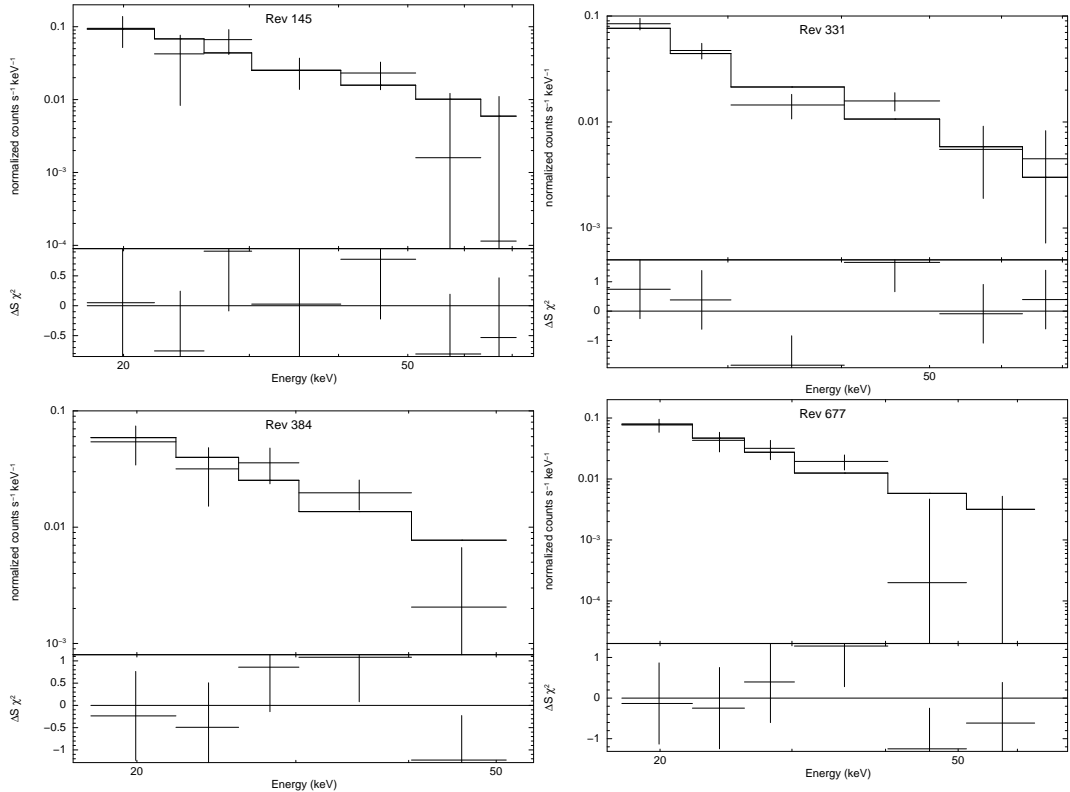

\centering
\includegraphics[angle=-90,width=7cm]{2s0114_spe_145.ps}
\includegraphics[angle=-90,width=7cm]{2s0114_spe_331.ps}
\includegraphics[angle=-90,width=7cm]{2s0114_spe_384.ps}
\includegraphics[angle=-90,width=7cm]{2s0114_spe_677.ps}
\caption{The hard X-ray spectrum of 2S 0114+65 obtained by
INTEGRAL/IBIS during its quiescent states: Rev. 145, 331, 384,
677. In quiescence, the source cannot be detected significantly
above $\sim 50$ keV. The spectra are fitted with a single pow-law
model. See the text for details. }
\end{figure*}

During the IBIS/ISGRI hard X-ray monitoring of the high mass X-ray
binary 2S 0114+65 from 2003 -- 2006, the source 2S 0114+65 was not
detected in some time intervals, or was detected with a low
significance level with a mean count rate of $<1$ cts s$^{-1}$.
During these time intervals, we have explained that 2S 0114+65 was
undergoing the quiescent states.

From Table 2, 2S 0114+65 was detected above 5$\sigma$ for four
revolutions in the quiescent states. The four spectra in these
revolutions are presented in Fig. 8. In the quiescent states, 2S
0114+65 cannot be detected significantly above 50 keV, and only a
few data points can be used in the spectral fitting. We also use a
single power-law model to fit all spectra, and spectral properties
are shown in Table 2. The flux and luminosity are given only in
the range of 20 -- 50 keV. In quiescence, X-ray luminosity from 20
-- 50 keV is around (2 -- 5)$\times 10^{35}$ erg s$^{-1}$. The
photon indexes distribute from 2.6 -- 3.2 with larger error bars,
so the hard X-ray spectral properties of 2S 0114+65 in quiescent
states may have no significant differences from those in active
states.

\subsection{Hard X-ray tails}

\begin{figure}
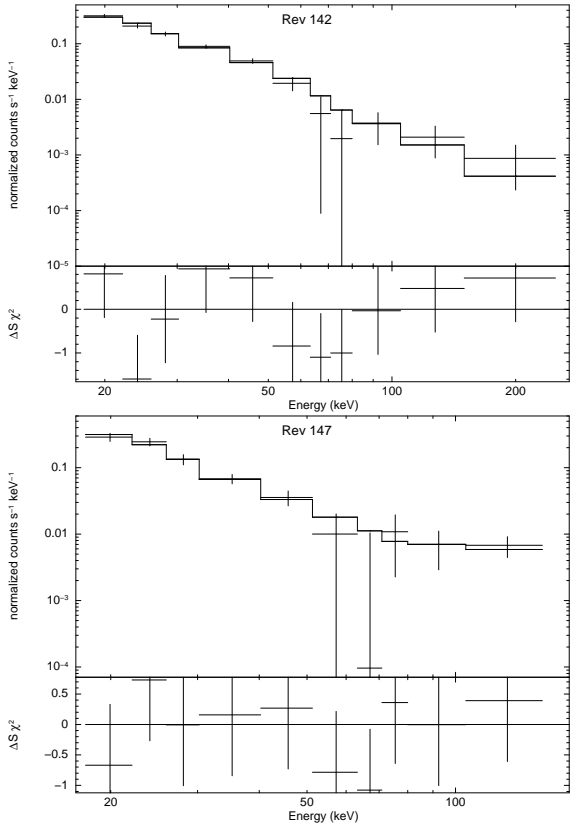

\centering
\includegraphics[angle=-90,width=7.5cm]{2s0114_spe_142.ps}
\includegraphics[angle=-90,width=7.5cm]{2s0114_spe_147.ps}
\caption{The hard X-ray spectra of 2S 0114+65 in two revolutions:
142 and 147. The hard X-ray tails above 70 keV are detected in the
spectra. See the text for the details.}
\end{figure}

\begin{table}

\caption{INTEGRAL/JEM-X observations of 2S 0114+65 in 18
revolutions when the mean off-axis angle is below $\sim 5^\circ$
in Table 1. The source flux in the range of 3 -- 35 keV in units
of cts/s and detection significance level are also shown.}

\begin{center}
\scriptsize
\begin{tabular}{l c l}

\hline \hline Rev. Num.  & Flux  & Detection significance   \\
\hline
238  & $<0.14$ & $<5\sigma$ \\
262 & $0.49\pm 0.11$  & $5\sigma$  \\
263 & $0.42\pm 0.10$  & $5\sigma$  \\
331 & $<0.12$  &  $<5\sigma$  \\
332 & $1.29\pm 0.06$  &  $24\sigma$ \\
333 & $0.61\pm 0.04$  &  $15\sigma$ \\
334 & $<0.15$  &  $<5\sigma$ \\
335 & $1.15\pm 0.07$  &  $16\sigma$ \\
454 & $<0.12$  &  $<5\sigma$ \\
528 & $<0.14$  &  $<5\sigma$ \\
664 & $1.65\pm 0.07$  &  $26\sigma$ \\
667 & $<0.25$  &  $<5\sigma$ \\
668 & $<0.23$  &  $<5\sigma$ \\
669 & $<0.18$  &  $<5\sigma$ \\
670 & $<0.18$  &  $<5\sigma$ \\
673 & $1.17\pm 0.06$  &  $20\sigma$ \\
675 & $0.79\pm 0.04$  &  $19\sigma$ \\
670 & $<0.09$  &  $<5\sigma$ \\
 \hline
\end{tabular}
\end{center}

\end{table}

den Hartog et al. (2006) reported the detection of the high energy
tail of 2S 0114+65 up to 120 keV without obvious high energy
cutoff with the IBIS data, in which the spectrum was depicted by a
combination of thermal bremsstrahlung and power-law models. This
hard X-ray tail was also indicated in the spectrum of 2S 0114+65
during the time interval of Rev 389 (see Fig. 7 \& Table 2). We
also check the IBIS data (Rev 142 -- 148 \& Rev 161 -- 162) used
by den Hartog et al. (2006). We also derived the hard X-ray
spectra of the source for each revolution when 2S 0114+65 was
bright. In Figure 9, we illustrate the spectra of 2S 0114+65 in
two revolutions: 142 and 147.

In Rev 142, the source was detected with a significance level of
$\sim 33\sigma$. The spectrum was fitted by a single power-law
model of $\Gamma\sim 2.6\pm 0.1$ (reduced $\chi^2\sim 2.049 ,\ 9
d.o.f.$) or a thermal bremsstrahlung model of $kT\sim 29.1\pm 2.7$
keV (reduced $\chi^2\sim 1.237 ,\ 9 d.o.f.$). The data points
above 80 keV cannot be fitted well. Thus we fit the spectrum using
a thermal bremsstrahlung model plus a single power-law model, with
$kT\sim 26.2\pm 4.2$ keV and $\Gamma\sim 1.1\pm 0.9$ (reduced
$\chi^2\sim 1.133,\ 7 d.o.f.$). Then the high energy tail added in
the model improved the fitting. The derived hard X-ray flux from
20 -- 100 keV was $\sim (2.2\pm 0.2) \times 10^{-10}$ erg
cm$^{-2}$ s$^{-1}$.

The hard X-ray spectrum for Rev 147 was fitted with a single
power-law model of $\Gamma\sim 2.7\pm 0.2$ (reduced $\chi^2\sim
1.476 ,\ 8 d.o.f.$) or a thermal bremsstrahlung model of $kT\sim
22.1\pm 4.1$ keV (reduced $\chi^2\sim 1.687 ,\ 8 d.o.f.$). The
simple model also cannot fit the spectrum well. We then fit the
spectrum using a thermal bremsstrahlung plus a single power-law
model, with $kT\sim 17.9\pm 3.1$ keV and $\Gamma\sim 0.7\pm 0.9$
(reduced $\chi^2\sim 0.899,\ 6 d.o.f.$). The derived X-ray flux in
the range of 20 -- 100 keV was $(1.9\pm 0.2)\times 10^{-10}$ erg
cm$^{-2}$ s$^{-1}$.

Thus, we confirmed the hard X-ray tails reported by den Hartog et
al. (2006) using the same datasets. We have detected the hard
X-ray tails in 2S 0114+65 at least in two observational
revolutions: 142, 147. Thus hard X-ray tails should really exist
in such accretion neutron star systems. For neutron star X-ray
binaries or accreting X-ray pulsars, their typical X-ray spectra
from 1 keV to tens of keV can be generally described by an
absorbed power-law model with a high energy exponential cut-off
(Hall et al. 2000). Thus detection of this high energy tail is
quite interesting for the accretion neutron star systems. The
spectra of neutron star X-ray binaries generally have no hard
X-ray tails which are different from those of black hole X-ray
binaries. Then it is very important to investigate the
characteristics and the origin of the hard X-ray tails in the
neutron star accretion binary 2S 0114+65?

\begin{figure}
\centering
\includegraphics[angle=0,width=8.5cm]{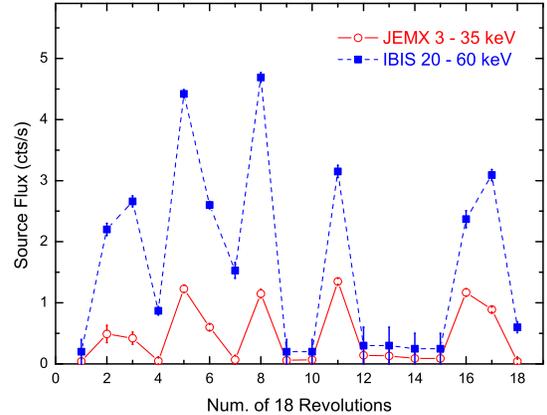}
\caption{The JEM-X flux (3 -- 35 keV) versus the IBIS/ISGRI flux
(20 -- 60 keV) for 2S 0114+65 in the 18 observational revolutions
listed in Table 3. }
\end{figure}

\begin{table*}

\caption{Spectral properties and searching for hard X-ray tails in
2S 0114+65 from different observations or work. Most of the
spectra were fitted by the model {\em abs*pow*highe}, some of them
by {\em pow*highe} or a single power-law model. The column density
is in units of $10^{22}$ cm$^{-2}$, $E_{\rm cut}$ and $E_{\rm
fold}$ in units of keV, flux in units of $10^{-10}$ erg cm$^{-2}$
s$^{-1}$ in the given energy range.}

\begin{center}
\scriptsize
\begin{tabular}{l c c c c c c c l}

\hline \hline Work & Instrument & $N_{\rm H}$ & $\Gamma$  & $E_{\rm cut}$  & $E_{\rm fold}$ & energy range (keV) & Flux  & hard X-ray tail? \\
\hline
Hall et al. (2000) & RXTE &$3.5\pm 0.3$ & $1.37\pm 0.05$ & $8.4\pm 0.4$ & $20.3\pm 1.2$ &  3 -- 20 & 1.09 & No \\
Bonning \& Falanga (2005)& INTEGRAL &- & 1.6  & 9 & 22.1 &   5 -- 100 & 2.4 & No \\
Masetti et al. (2006)& BeppoSAX & $9.7\pm 0.9$ & $1.33\pm 0.16$  & $12\pm 3$ & $21\pm 4$ & 1.5 -- 100 & 3.3 & No \\
den Hartog et al. (2006) & INTEGRAL & - & $3.1\pm 0.06$ & - & - & 20 -- 150 &  1.7 & Yes \\
Farrell et al. (2008)& RXTE & $3.2\pm 0.9$ & $1.1\pm 0.1$  & 6.0$\pm 0.7$ & 15$\pm 3$ & 3 -- 50  & 2.3 & No \\
\hline
{\bf This work} &  & &   & &  &   &  \\
\hline  Rev 332 & INTEGRAL & $6.9\pm 2.8$ & $1.2\pm 0.1$ & $6.1\pm 2.7$& $28.5\pm 3.2$& 3 -- 100 & 6.4 & No \\
Rev 333 & INTEGRAL &- & $0.9\pm 0.1$ & $6.5\pm 2.1$ & $20.5\pm 2.1$ & 3 -- 100 & 3.4 & Yes \\
Rev 335 & INTEGRAL &- & $0.7\pm 0.1$ & $5.9\pm 2.2$ & $19.4\pm 1.7$ & 3 -- 100 & 5.9 & ? \\
Rev 664 & INTEGRAL & - & $0.7\pm 0.1$ & $8.3\pm 0.5$ & $8.4\pm
0.4$ & 3 -- 100 & 7.0 & Yes \\
Rev 673 & INTEGRAL & - & 1.2$\pm 0.2$ & $7.7\pm 1.4$ &$13.6\pm 1.7$ & 3 -- 100 & 4.9 & Yes \\
Rev 675 &INTEGRAL & $6.5\pm 2.7$ & $1.1\pm 0.2$ & $6.9\pm 2.6$ & $25.2\pm 3.9$& 3 -- 100  & 4.6 & No \\
\hline
\end{tabular}
\end{center}

\end{table*}

To constrain the hard X-ray tails in the spectra of 2S 0114+65, we
will try to combine the soft X-ray band data in the range of 3 --
35 keV from JEM-X aboard INTEGRAL with the IBIS data above 18 keV.
JEM-X has a small field of view so we can study 2S 0114+65 using
JEM-X only when the mean off-axis angle on the source is below
5$^\circ$. According to Table 1, only 18 revolutions meet the
above detection condition. The JEM-X detection results on 2S
0114+65 for these 18 revolution are shown in Table 3. 2S 0114+65
was detected significantly by JEM-X (e.g., $> \sim 15\sigma$ in
the energy band of 3 -- 35 keV) in six revolutions. In Fig. 10, we
plot the variations of source flux detected both by JEM-X and IBIS
in the 18 revolutions. JEM-X can detect 2S 0114+65 when IBIS has
very high detection significance level ($>30\sigma$, except for
Rev 673, $\sim 20\sigma$ for JEM-X and $\sim 17\sigma$ for IBIS).

We would try to study the combined spectral properties with JEM-X
and IBIS, so we select the observational revolutions when 2S
0114+65 was detected with significance level $>\sim 15\sigma$ both
by JEM-X and IBIS. Six revolutions are used for the combined
spectral analysis: Rev. 332, 333, 335, 664, 673 and 675 (see Table
3). The cross-calibration studies on the JEM-X and IBIS/ISGRI
detectors have been done by Jourdain et al. (2008) using the Crab
observation data. The calibration between JEM-X and IBIS/ISGRI for
OSA 7.0 can be good enough within $\sim 6\%$. JEM-X only gave the
upper limits for the weak source 2S 0114+65 above 20 keV in our
observations, so we only use the spectral data points from 3 keV
up to $\sim 25$ keV in the spectral analysis.

With the JEM-X data combined with the IBIS data, we can construct
the X-ray spectrum from 3 -- 100 keV even up to 200 keV for 2S
0114+65 to study the properties of the hard X-ray tail. In Fig.
11, we present the X-ray spectra from 3 -- 100 keV (up to 200 keV
in some cases) of 2S 0114+65 in six observational revolutions when
both JEM-X and IBIS had the significant detections: Revs. 332,
333, 335, 664, 673 and 675. All the spectra are first fitted with
an absorbed power-law model plus high energy exponential cutoff
(hereafter {\rm abs*pow*highe}). In two Revs. 332 and 675, the
high column densities ($\sim (6-7)\times 10^{22}$ cm$^{-2}$) are
derived, and the {\rm abs*pow*highe} model fit the spectra well up
to 120 keV, no hard X-ray tails are detected. In other Revs,
spectral fittings with the {\rm abs*pow*highe} model imply a low
column density (near zero in fittings or an upper limit of $\sim
3\times 10^{22}$ cm$^{-2}$), so these spectra can be directly
fitted by a power-law plus high energy exponential cutoff
(hereafter {\em pow*highe}). In three Revs. 333, 664 and 675, the
significant flux excesses are detected above 70 keV, suggesting
the existence of hard X-ray tails. In Rev 335, only one data point
above 70 keV appears to be an excess, so we cannot conclude that
it could be the hard X-ray tail.

From the spectral analysis in the energy band of 3 -- 100 keV even
up to 200 keV (see Fig. 11 and Table 4), the hard X-ray tails
really exist in 2S 0114+65 for some time intervals. Our results
confirmed den Hartog's studies, but most previous studies have not
detected the hard X-ray tails. For comparison, we collected
previous spectral measurements in Table 4. The work by Hall et al.
(2000) and Farrell et al. (2008) used the data covering the energy
range below 50 keV, so it is reasonable that they cannot detect
the hard X-ray tails generally above 60 keV. Masetti et al. (2006)
used the BeppoSAX data in the range of 1.5 -- 100 keV, and did not
detect the high energy tail but derived a large column density of
$9.7\times 10^{22}$ cm$^{-2}$. Their result is still consistent
with our results: in two revolutions 332 and 675, no hard X-ray
tails are detected with large column densities (see Table 4).

\begin{figure*}
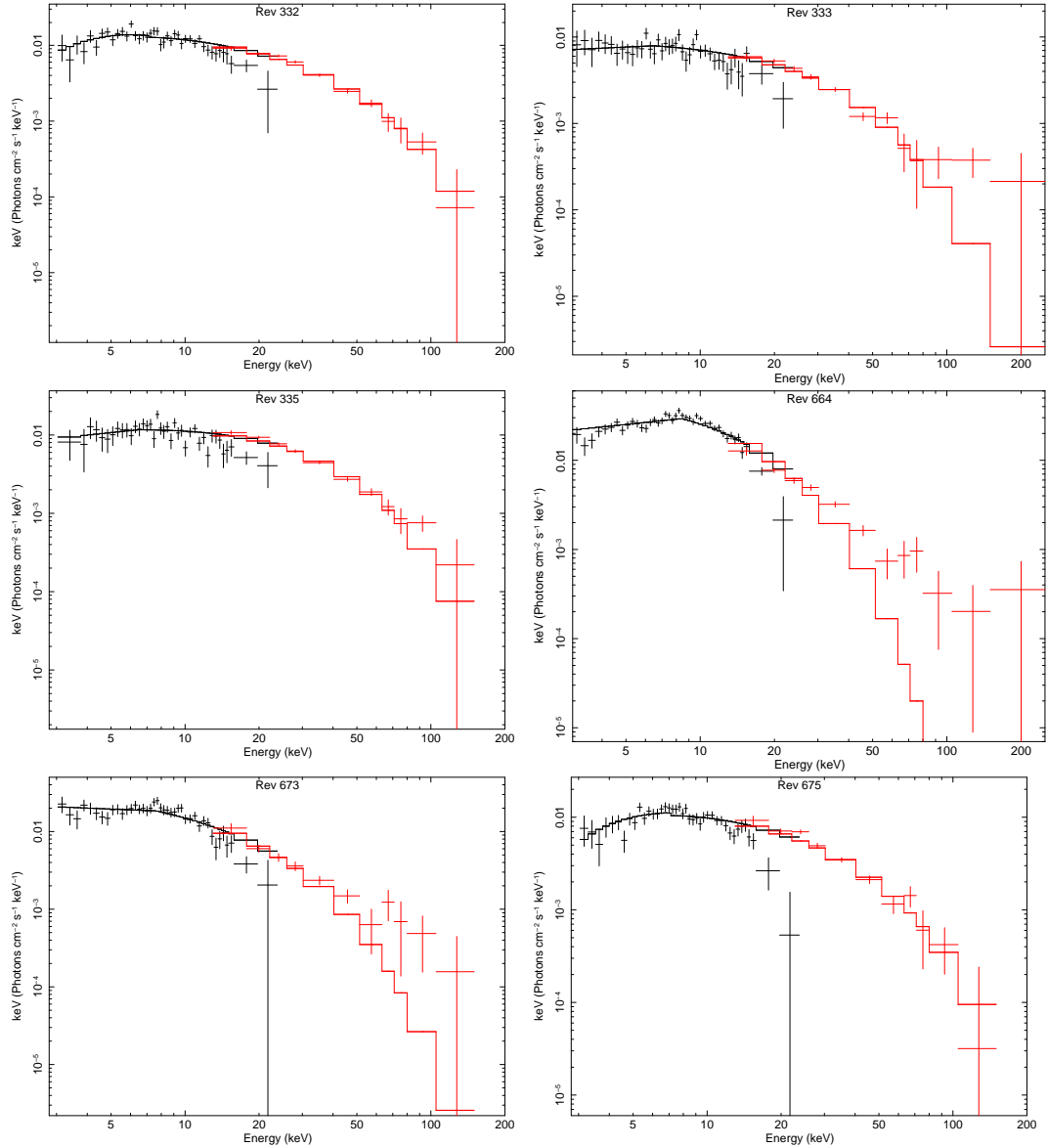

\centering
\includegraphics[angle=-90,width=7.0cm]{2s0114_300c_332euf.ps}
\includegraphics[angle=-90,width=7.0cm]{2s0114_300c_333euf.ps}
\includegraphics[angle=-90,width=7.0cm]{2s0114_300c_335euf.ps}
\includegraphics[angle=-90,width=7.0cm]{2s0114_300c_664euf.ps}
\includegraphics[angle=-90,width=7.0cm]{2s0114_300c_673euf.ps}
\includegraphics[angle=-90,width=7.0cm]{2s0114_300c_675euf.ps}
\caption{The unfolded hard X-ray spectra from 3 -- 100 keV of 2S
0114+65 with the combined observational data of IBIS and JEM-X in
six time intervals: Revs. 332, 333, 335, 664, 673 and 675. }
\end{figure*}

Therefore, we find that the detection of hard X-ray tails is
sensitive to the column density. The hard X-ray tails are detected
when the column density is very low (near zero in fittings). The
high column density may lead to the disappearance of the hard
X-ray tails. It is quite interesting because generally the large
column density induces strong absorption in soft X-ray bands of
0.1 -- 5 keV. Thus, our results may provide a clue to probe the
origin of the hard X-ray tails in neutron star accretion systems,
especially the relevant information for the physical environment.

\subsection{Orbital phase-resolved spectral properties}

\begin{figure*}
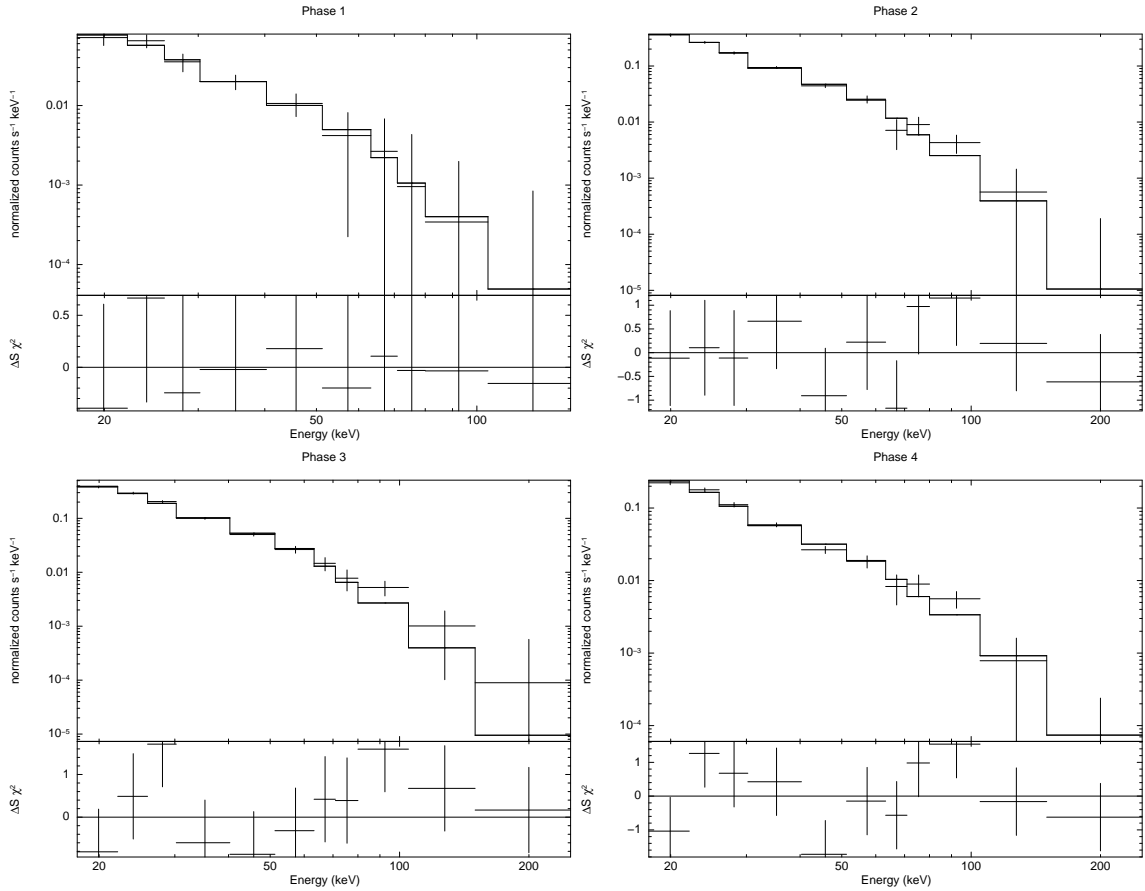

\centering
\includegraphics[angle=-90,width=7.5cm]{2s0114_pha1.ps}
\includegraphics[angle=-90,width=7.5cm]{2s0114_pha2.ps}
\includegraphics[angle=-90,width=7.5cm]{2s0114_pha3.ps}
\includegraphics[angle=-90,width=7.5cm]{2s0114_pha4.ps}
\caption{The hard X-ray spectra of 2S 0114+65 in four orbital
phases: phase 1 (0.0--0.25); phase 2 (0.25 -- 0.5); phase 3
(0.5--0.75); and phase 4 (0.75 -- 1.0). For a comparison, all
spectra are fitted with the same spectral model: a power-law model
with a high energy cutoff. See details in the text. }
\end{figure*}

The high mass X-ray binary 2S 0114+65 has an orbital period of
$\sim 11.59$ day (Crampton et al. 1985; Corbet et al. 1999). The
X-ray light curves of 2S 0114+65 shown in Fig. 2 cover about 3
orbital periods (12 continuous observational revolutions from Rev.
384 -- Rev. 395 by IBIS/ISGRI). Then we try to derive the hard
X-ray spectra for different orbital phases with the IBIS
observational data from Rev. 384 -- Rev. 395. To obtain the good
detection significance for the spectral analysis in each orbital
phase, we divide the orbital phase into four parts, and obtain the
spectral properties for each orbital phase to show the spectral
variations in different orbital phases. The phase-resolved data
point starts from Rev. 384.

We derived four spectra of four different orbital phases (Figure
12). For a comparison, we used the same model to fit the continuum
of all the spectra from 18 -- 150 keV: a power-law model with a
high energy exponential cutoff. We also compare variations of the
three spectral parameters over orbital phases: photon index
$\Gamma$, the cutoff energy $E_{\rm cut}$ and the e-folding energy
$E_{\rm fold}$ for four orbital phases.

The hard X-ray spectral parameters of 2S 0114+65 in the different
orbital phases were presented together for comparison in Table 5.
In phase 1, the photon index was $\Gamma\sim 2.56\pm 0.41$, and
$E_{\rm cut}\sim 10.8 \pm 2.1$ keV with $E_{\rm fold}\sim 19.3\pm
4.2$ keV (reduced $\chi^2\sim 0.968 ,\ 6 d.o.f.$), and above 60
keV, 2S 0114+65 can not be detected significantly during phase 1.
In phase 2, $\Gamma\sim 1.53\pm 0.35$, $E_{\rm cut}\sim 18.9 \pm
1.6$ keV with $E_{\rm fold}\sim 27.4\pm 4.8$ keV (reduced
$\chi^2\sim 0.717 ,\ 7 d.o.f.$); in phase 3, $\Gamma\sim 1.92\pm
0.31$, $E_{\rm cut}\sim 25.8 \pm 1.3$ keV with $E_{\rm fold}\sim
35.9\pm 3.1$ keV (reduced $\chi^2\sim 0.799 ,\ 7 d.o.f.$); and in
phase 4, $\Gamma\sim 2.18\pm 0.38$, and the high energy cutoff
given as $E_{\rm cut}\sim 24.0 \pm 2.1$ keV with $E_{\rm fold}\sim
69.8\pm 10.2$ keV (reduced $\chi^2\sim 1.18 ,\ 7 d.o.f.$).

From the continuum fits, we found that the spectral properties of
2S 0114+65 changed with the orbital phases. In Figure 13, we plot
the changes of the three fitted parameters: $\Gamma$, $E_{\rm
cut}$ and $E_{\rm fold}$ with the orbital phases. The changes of
the photon index and $E_{\rm cut}$ looks like a sinusoidal curve:
at the maximum of the light curve, the spectrum is harder with
smaller photon index with a larger high energy cut-off. The
variation of the power-law photon index over orbital phase
anticorrelates with hard X-ray flux, and the variation of $E_{\rm
cut}$ has a positive correlation with the hard X-ray flux. Farrell
et al. (2008) found that the variations of neutral hydrogen column
density and power-law photon index also have anti-correlations
with the X-ray flux using the RXTE/PCA 2 -- 9 keV data. At the
minimum of the light curve, the increase of the column density and
photon index suggested the presence of the strong absorption
effect by the dense stellar wind. Thus, the change of the hard
X-ray properties in 2S 0114+65 over orbital phase is still
consistent with those in soft X-ray bands.

\begin{figure}
\centering
\includegraphics[angle=0,width=8.cm]{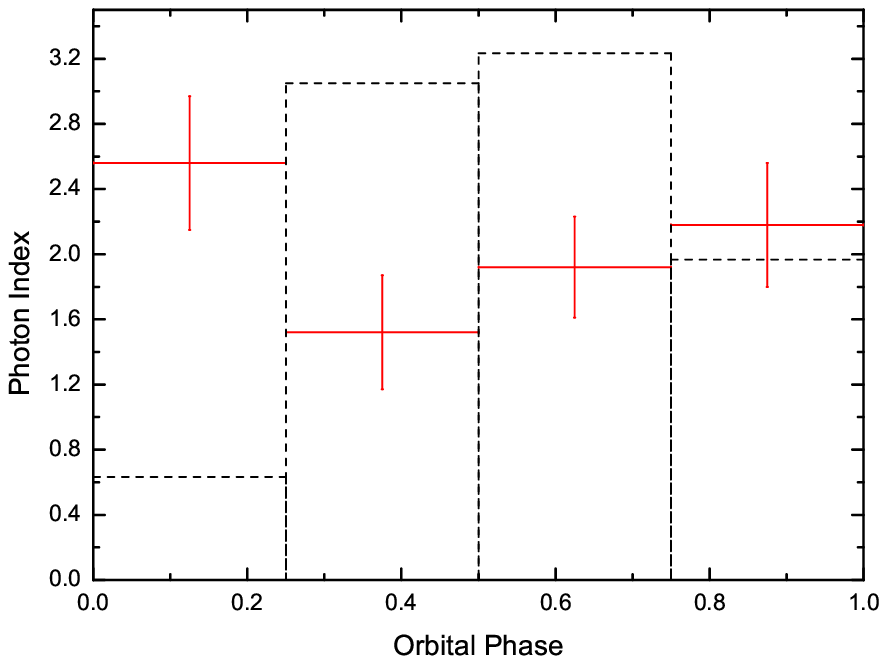}
\includegraphics[angle=0,width=8.cm]{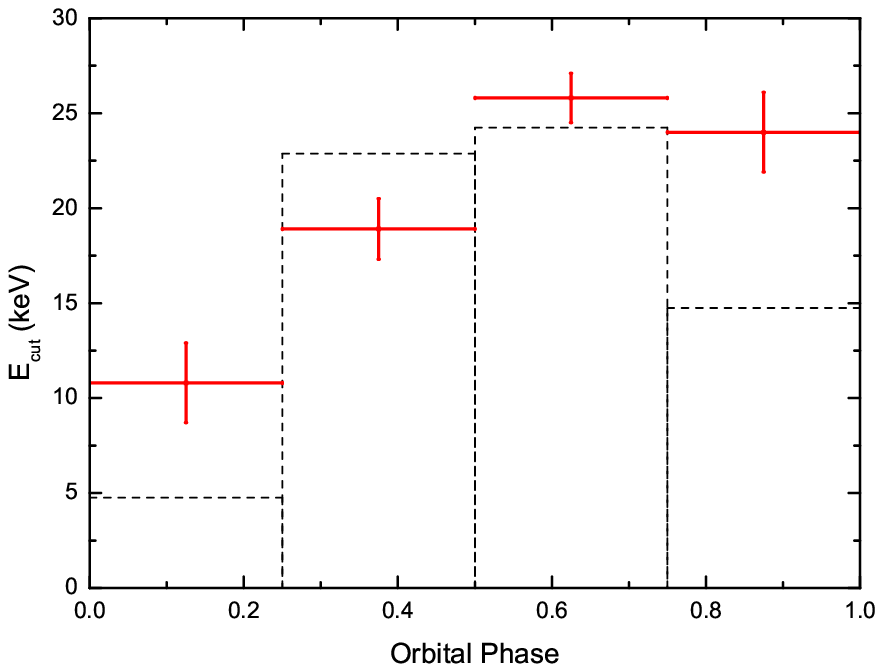}
\includegraphics[angle=0,width=8.cm]{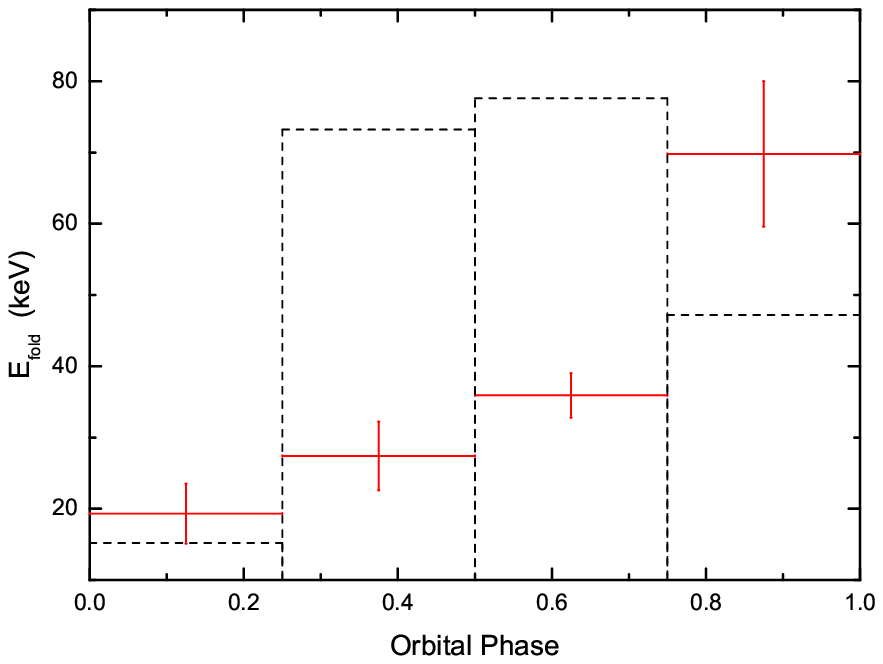}
\caption{The spectral properties of 2S 0114+65 with the different
orbital phases. For four spectra of four orbital phases, we use
the power-law model with a high energy cutoff to fit the continuum
of all the spectra. And we show the fitted parameters: the photon
index (top), the high cutoff energy ($E_{\rm cut}$, middle) and
the exponential folding energy ($E_{\rm fold}$, bottom) over four
orbital phases. The orbital variability of the normalized flux
rates in hard X-ray bands is also plotted (dashed curves) for a
comparison }
\end{figure}

In addition, we also tried to search for the possible cyclotron
absorption features after the continuum fittings of the four
spectra (see the residuals in Figure 12). However, no significant
absorption features were detected in all orbital phase-resolved
spectra.

\section{Summary and discussion}

\begin{table}

\caption{Hard X-ray spectral properties of 2S 0114+65 in different
orbital phases. All spectra are fitted by the pow*highe model. }

\begin{center}
\scriptsize
\begin{tabular}{l c c l}

\hline \hline Phase  & $\Gamma$  & $E_{\rm cut}$ (keV) & $E_{\rm fold}$ (keV)   \\
\hline
0.0 -- 0.25  & $2.56\pm 0.41$ & $10.8\pm 2.1$ & 19.3$\pm 4.2$ \\
0.25 -- 0.5 & $1.53\pm 0.35$  & $18.9\pm 1.6$ & $27.4\pm 4.8$  \\
0.5 -- 0.75& $1.92\pm 0.31$  & $25.8\pm 1.3$ &$35.9\pm 3.1$  \\
0.75 -- 1.0  & $2.18\pm 0.38$  & 24.0$\pm 2.1$ & 69.8$\pm 10.2$   \\
\hline
\end{tabular}
\end{center}

\end{table}

\begin{figure}
\centering
\includegraphics[angle=0,width=9cm]{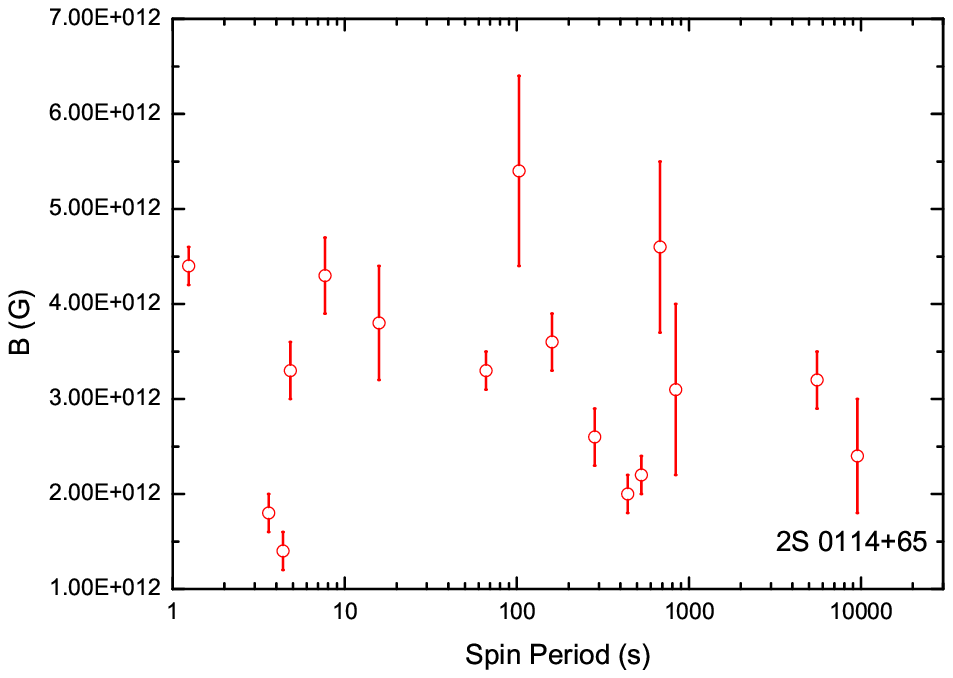}
\caption{Magnetic field of X-ray accretion pulsars from
measurements of cyclotron resonant absorption features. Spin
period of the neutron stars are also shown. For 2S 0114+65, the
possible value of magnetic field are displayed by assuming 22 keV
(Bonning \& Falanga 2005) as the fundamental line energy. The
other points are collected from Coburn et al. (2002), Kreykenbohm
et al. (2005), McBride et al. (2006), and Wang (2009).}
\end{figure}

In this paper, we have carried out the long-term hard X-ray
monitoring of the high mass X-ray binary 2S 0114+65 with the soft
gamma-ray detector IBIS/ISGRI aboard INTEGRAL. 2S 0114+65 is a
variable X-ray source whose luminosity changes with the orbital
phase. The hard X-ray properties changed in different accretion
states and over the orbital phases. The characteristics of the
spin period of the neutron star in 2S 0114+65, the variations of
hard X-ray properties over orbital phases and conspicuous features
of the hard X-ray tails are investigated in detail.

The main results from our hard X-ray monitoring project on 2S 0114+65 are presented as follows: \\
(1) The spin evolution of the neutron star in 2S 0114+65 is
studied. With the observations covering near five years from 2003
December to 2008 May, we obtained the values of the spin period of
the neutron star in 2S 0114+65 in six time intervals when 2S
0114+65 was detected with a high significance level: 2003 Dec,
$P=9612\pm 20$ s ; 2004 Feb, $P=9600\pm 20$ s ; 2004 Dec,
$P=9570\pm 20$ s ; 2005 July, $P=9555\pm 15$ s; 2005 Dec -- 2006
Jan, $P=9520\pm 20$ s; 2008 April $P=9475\pm 25$ s. Thus we
obtained a spin-up rate of the neutron star in 2S 0114+65 $\sim
1.09\times 10^{-6}$ s s$^{-1}$ from 2003 -- 2008. Compared with
the previous reported spin-up rate, the neutron star in 2S 0114+65
was spinning up, and the spin-up rate was accelerating. \\
(2) Hard X-ray spectral properties of 2S 0114+65 changed in
different time intervals and accretion states (see Figs. 7 \& 8).
The detected hard X-ray luminosity from 20 -- 100 keV varies from
$10^{35}$ erg to $4\times 10^{36}$ erg. For the comparison, the
spectra of 2S 0114+65 in different states are fitted with the same
spectral model -- a single power-law model. In active states, the
photon index varies from $2.5 - 2.8$; in quiescent states, 2S
0114+65 was not detected significantly above 50 keV with
$\Gamma\sim 2.6 -3.2$. The cyclotron absorption line features at
22 keV and 44 keV cannot be confirmed by our spectral
studies.\\
(3) Hard X-ray tails above 60 keV are detected in the spectra of
2S 0114+65 in some revolutions. The 20 -- 120 keV spectra can be
fitted by the bremsstrahlung model plus a power-law (see Fig. 9).
We also obtained the broadband spectra from 3 -- 100 keV using
both JEM-X and IBIS data to constrain the continuum spectral
properties and study the characteristics of hard X-ray tails. The
3 -- 100 keV can generally be described by an absorbed power-law
model with high energy cutoff. When the derived values of column
density are higher than $\sim 3\times 10^{22}$ cm$^{-2}$, no hard
X-ray tails are detected. While in the case of low column density
(column density is near zero in our fittings), hard X-ray tails
are detected. The flux excesses above 60 keV cannot be fitted with
the model {\em abs*pow*highe}. Therefore, detections of hard X-ray
tails are sensitive to the column density. High column density may
lead to the disappearance of the hard X-ray tails in the spectra
of 2S 0114+65. Our results confirm the den Hartog's report on hard
X-ray tail in 2S 0114+65, and can also reasonably explain why
previous studies did not detected high energy tails (see Table 4).
Though the origin of hard X-ray tails in neutron star accretion
systems such as 2S 0114+65 is unclear,
our results would provide a good clue to probe the origin.\\
(4) Finally, we also study the variations of spectral properties
of 2S 0114+65 with different orbital phases. All hard X-ray
spectra from 20 -- 150 keV were fitted with a power-law model with
a high energy cutoff. The variation of the power-law photon index
over orbital phase anticorrelates with hard X-ray flux, and the
variation of $E_{\rm cut}$ has a positive correlation with the
hard X-ray flux, suggesting that the harder spectrum at the
maximum of the light curve. The variations of spectral properties
of 2S 0114+65 over orbital phase are consistent with the feature
in the highly obscured binary system. The possible cyclotron
absorption features around 22 keV and 44 keV were not detected in
all orbital phase-resolved spectra.

\begin{figure}
\centering
\includegraphics[angle=0,width=8.5cm]{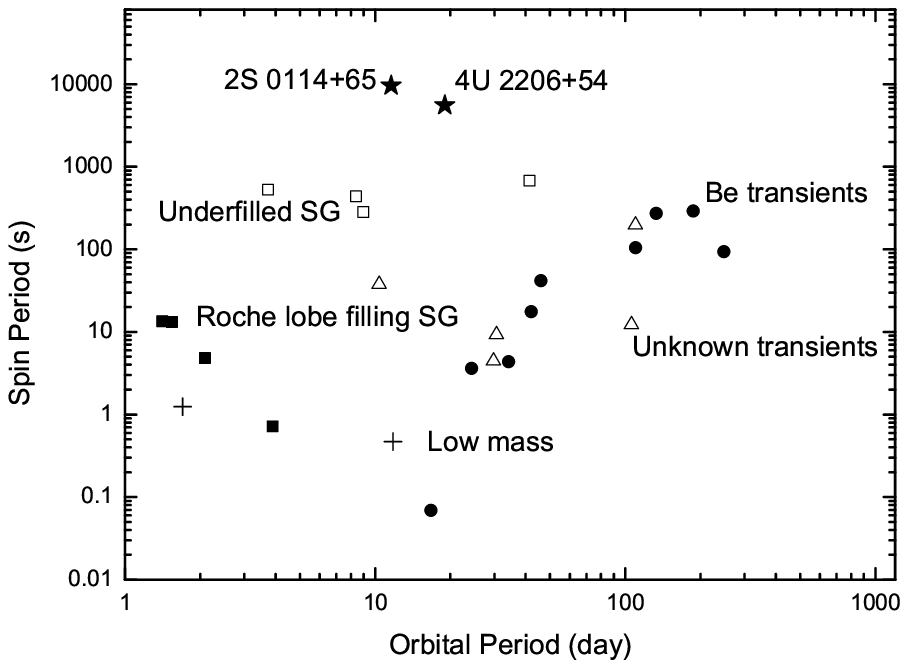}
\caption{The spin period - orbital period diagram for the
accreting neutron star systems (the Corbet diagram, Corbet 1986).
The data point for 2S 0114+65 is taken from this work and Farrell
et al. (2008); the data point for 4U 2206+54 is taken from Wang
(2009); other points are taken from Bildsten et al. (1997). There
exist a positive correlation between $P_{\rm spin}- P_{\rm orbit}$
for the Be transient systems; and a possible negative correlation
for the disk-accretion system including Roche Lobe Filling
Supergiants and low mass systems; For the underfilled Roche Lobe
Supergiants and two long-pulsation systems 2S 0014+65, 4U 2206+54
which all belong to wind-fed accretion systems, the possible
relation between $P_{\rm spin}- P_{\rm orbit}$ is different from
that of the Be transients. }
\end{figure}

The possible electron cyclotron resonant scattering features
(CRSF) around 22 keV (Bonning \& Falanga 2005) was not detected in
our long-term INTEGRAL/IBIS observations. The CRSF around 22 keV
was also not detected by other independent observations (Masetti
et al. 2006; den Hartog et al. 2006; Farrell et al. 2008). The
possible absorption feature at $\sim 44$ keV was also not evident
in our spectra but was only hinted in one observational revolution
in active states. In theories, the scattering cross-section is
dependent on the angle between photon direction and the magnetic
field (Schoenherr et al. 2007), so CRSFs provide a diagnostic of
the emission geometry and the physical conditions of the radiating
plasma (Harding \& Lai 2006). For example, when the photon
direction and magnetic field vector are aligned, no CRSFs are
expected. Variations in the CRSFs are also expected with the
viewing geometry, accretion geometry or electron density and even
the accretion rate (see Araya \& Harding 1999, 2000). The
relativistic cyclotron absorption line energy would have
non-harmonic line spacing and changed with luminosity and some
relevant parameters (Araya \& Harding 1999; Klochkov et al. 2008).
Thus possible CRSFs are not excluded. If we assume the 22 keV line
as the fundamental, the magnetized neutron star with the magnetic
field of $\sim 2.5\times 10^{12}$ G is harboured in 2S 0114+65
which is preferred by the formation scenario proposed by Li \& van
den Heuvel (1999). Up to now, the cyclotron absorption line
features are detected in more than 10 high mass X-ray binary
systems. The energy of the fundamental line distributes from 10 --
50 keV, corresponding to a magnetic field of $\sim (1.6-6)\times
10^{12}$ G (see Figure 14). In Figure 13, we show the magnetic
field of neutron stars derived from the electron cyclotron
absorption line measurements versus their spin periods in
accretion X-ray binaries up to now; for 2S 0114+65, the possible
CRSF at 22 keV is taken. We found that the magnetic field has no
correlation to the spin period of the neutron stars in high mass
X-ray binaries.

We have studied the hard X-ray spectral properties of 2S 0114+65
in different accretion states. The change of the spectral
properties of 2S 0114+65 with the orbital phase in hard X-rays (18
-- 150 keV) were presented. In soft X-rays (e.g., 2 -- 9 keV), the
spectral properties also changed with the orbital phases as
suggested by the RXTE/PCA observations (Farrell et al. 2008). In
soft X-ray bands, the observed variability of the neutral hydrogen
column density over the orbital period indicated that the variable
absorption by the dense stellar wind medium is the mechanism
behind the orbital modulation (Farrell et al. 2008). The
variability of fitted photon index from 2 -- 9 keV against orbital
phase showed a good anti-correlation between the photon index and
count rate, implying a harder spectrum around the maximum of the
count. In hard X-ray bands, the variability of photon index over
orbital phase has an anticorrelation with the flux. In addition,
we also derived the variation of high energy cut-off over orbital
phase which has a positive correlation with the flux. Thus around
the maximum of the count rate the spectrum appeared harder, which
is similar to the case in soft X-rays.

The high energy tails in the spectra of 2S 0114+65 above 70 keV
are detected and studied in details using the IBIS data combined
with JEM-X observations. High column density would lead to
disappearance of hard X-ray tails. The hard X-ray tails in X-ray
binaries have been reported in black hole accretion systems and
low-mass X-ray binaries. How to produce the hard X-ray tails above
70 keV for accreting neutron star in high mass X-ray binaries
especially in the wind-fed accretion systems is unclear.
Generally, hard X-ray tails are suggested to be linked to three
possible scenarios: compact jets, accretion disc or hot corona.
The results from radio observations of 2S 0114+65 (Farrell et al.
2007) are against the possibility of the presence of jets. In
addition, the stable accretion disc seems to be impossible to form
in stellar-wind accretion systems like 2S 0114+65, but the
transient accretion disc is still possible (Crampton et al. 1985).
Hot corona with temperature of $kT> 10$ keV around inner part of
stable accretion disc is suggested in X-ray black hole binary
systems. Formation of hot corona is not so clear yet but it may be
related to outflows from accretion compact objects. So it is
possible that hot corona exists near neutron stars for wind-fed
accretion systems like 2S 014+65. The formation of the corona and
production processes of hard X-ray tails are beyond the scope of
this paper. Moreover, from our observations, the dense accretion
materials or strong winds prevent the formation of hot corona or
depress the comptonization effects. In summary, hard X-ray tails
in accretion neutron star systems like 2S 0114+65 is interesting
for both theorists and observers, requiring future work on both
theories and observations.

2S 0114+65 is one of the slowest pulsation neutron star systems.
In the $P_{\rm spin}-P_{\rm orbit}$ diagram (see Fig. 15), 2S
0114+65 and the other super-slow pulsation neutron star binary 4U
2206+54 ($P_{\rm spin}\sim 1.57$ hr, $P_{\rm orb}\sim 19.12$ days,
see Corbet et al. 2007; Reig et al. 2009; Wang 2009) have the
similar properties to underfilled Roche Lobe Supergiants which are
also powered by direct wind accretion. These systems may follow a
$P_{\rm spin}-P_{\rm orbit}$ relation quite different from that of
the Be transient systems (Fig. 15). 2S 0114+65 and 4U 2206+54 are
the only two known super-slow pulsation neutron star high mass
X-ray binaries ($P_{\rm spin}> 1000$ s), and two possible
super-slow X-ray pulsar candidates were also reported recently: 1E
161348-5055 in a young supernova remnant RCW 103 ($P_{\rm
spin}\sim 6.67$ hr, De Luca et al. 2006), and a wind-accretion
symbiotic low mass X-ray binary 4U 1954+319 ($P_{\rm spin}\sim 5$
hr, Mattana et al. 2006). The formation mechanisms of these
super-slow X-ray pulsars are unclear. Li \& van den Heuvel (1999)
have studied the origin of the long pulsation period X-ray pulsars
and suggested that a slow period is possible if the neutron star
was born as a magnetar with an initial magnetic field $\geq
10^{14}$ G, decaying to a current value of 10$^{12}$ G, allowing
the neutron star to spin down to the measured spin period within
the lifetime (Myr) of the companion. An alternative formation
channel proposed by Ikhsanov (2007) showed that an initial
magnetic field strength of $B\gg 10^{13}$ G is not necessary if
the evolutionary sequence of the neutron star consisted of both
supersonic and subsonic propeller phases. Anyway, both scenarios
predicted that the equilibrium period for 2S 0114+65 would be less
than 26 min, and this spin period will be approached on time
scales of $<100$ yr for disc accretion and $<1000$ yr for stellar
wind accretion (Ikhsanov 2007). From our measurements, the spin-up
rate of 2S 0114+65 at present is around $1.1\times 10^{-6}$ s
s$^{-1}$, and the spin-up rate was still accelerating in the last
30 years. If 2S 0114+65 would still spin up in the future, the
pulsar in 2S 0114+65 would reach its equilibrium period less than
200 years. The very short time scale suggested that the long
pulsation period X-ray pulsars should be very rare. Presently it
is fortunate (maybe strange from the probability of detection)
that at least three candidates are discovered. However, it is
still possible that 2S 0114+65 and other super-slow X-ray pulsars
are still young systems. Thus the current understanding of the
pulsar formation in different systems is quite deficient. Further
theoretic work may be required. It is hoped that future
observations of the spin-period evolution of 2S 0114+65 and other
similar systems would help us to understand the formation
mechanism of the long spin period.

\section*{Acknowledgments}
We are very grateful to the referee's fruitful comments and
suggestions to improve the paper. This paper is based on
observations of INTEGRAL, an ESA project with instrument and
science data centre funded by ESA member states (principle
investigator countries: Demark, France, Germany, Italy,
Switzerland and Spain), the Czech Republic and Poland, and with
participation of Russia and US. W. Wang is supported by the
National Natural Science Foundation of China under grants
10803009, 10833003, 11073030.

\appendix

\onecolumn

In the {\bf Appendix}, we show an extended table (compared with
Table 2 in the paper)to present the spectral fittings with
different models, like a single power law model {\em power}, a
thermal bremsstrahlung model {\em bremss}, a power-law plus
exponential high energy cutoff {\em pow*highe}. For some
revolutions, the spectra show the possible high energy tails, we
would fit the spectra with the combined model: the bremsstrahlung
plus a power-law model. Here we just display the spectral
parameters for each revolution with different model fittings, let
the readers independently justify which model would be more
acceptable to fit the spectrum of 2S 0114+65 in different
accretion states. The hard X-ray fluxes in the range of 20 -- 100
keV for the active states and 20 -- 50 keV for quiescent states in
different fits are presented.

\begin{table*}

\caption{Spectral properties of 2S 0114+65 in different accretion
states and datasets (for observations in each revolution).}

\begin{center}
\scriptsize
\begin{tabular}{l c c c c c l}

\hline \hline Rev. Num. & Model & $\Gamma$ / $kT$ (keV) & $E_{\rm cut}$ (keV) & $E_{\rm fold}$ (keV) &  Flux ($10^{-10}$ erg cm$^{-2}$ s$^{-1}$) & reduced $\chi^2$ \\
\hline
{\bf The active states} &  &   &  &  &   &  \\
\hline
142 & power & $2.6\pm 0.1$ & - & - &  $2.9\pm 0.2$ & $2.049 (9 d.o.f.)$ \\
 & bremss & 29.1$\pm 2.7$ & - & - &  $2.2\pm 0.2$ & $1.237 (9 d.o.f.)$ \\
  & bremss+power & 1.1$\pm 0.9$/26.2$\pm 4.2$ & - & -& $2.2\pm 0.2$& $1.133 (7 d.o.f)$ \\
161  & power & $2.8\pm 0.1$ & - & - &  $2.3\pm 0.4$ &  $1.252 (25 d.o.f.)$ \\
  & bremss & $22.8\pm 1.6$ & - & - &  $2.1\pm 0.4$ & $0.791 (25 d.o.f.)$ \\
  & pow*highe & $2.0\pm 0.8$ & 24.1$\pm 8.2$ & 33.0$\pm 16.9$ &$2.1\pm 0.4$ & $0.810 (23 d.o.f.)$  \\
262  & power & $2.6\pm 0.2$   &  - & - &  $1.6\pm 0.3$ &  $0.847 (26 d.o.f.)$ \\
 & bremss & $27.9\pm 3.9$ &  - & - &  $1.5\pm 0.3$ &  $0.704 (26 d.o.f.)$ \\
 & pow*highe & $1.7\pm 1.1$  & $22.4\pm 11.7$ & $35.7\pm 30.8$ &   $1.5\pm 0.3$ & $0.759 (24 d.o.f.)$ \\
263  & power & $2.7\pm 0.1$   &  - & - &   $1.9\pm 0.3$ &  $1.177 (26 d.o.f.)$ \\
 & bremss & $26.8\pm 2.3$   &  - & - &   $1.7\pm 0.3$ &  $1.288 (26 d.o.f.)$ \\
 & pow*highe & $2.1\pm 0.9$  & $23.3\pm 17.1$ & $50.7\pm 45.8$ &   $1.8\pm 0.3$ & $1.244 (24 d.o.f.)$ \\
264 & pow & $2.7\pm 0.1$  & - & - &  $2.5\pm 0.4$ & $0.793 (24 d.o.f.)$ \\
 & bremss & $26.5\pm 2.5$  & - & - &  $2.4\pm 0.4$ & $0.705 (24 d.o.f.)$ \\
 & pow*highe & $1.7\pm 0.7$  & 15.2$\pm 14.5$ & 33.4$\pm 21.6$ &   $2.4\pm 0.4$ & $0.729 (22 d.o.f.)$ \\
265 & pow & $2.6\pm 0.1$  & - & - &  $3.9\pm 0.3$ & $1.253 (27 d.o.f.)$ \\
 & bremss & $26.8\pm 1.5$  & - & - &  $3.6\pm 0.3$ & $0.975 (27 d.o.f.)$ \\
& pow*highe & $2.1\pm 0.3$  & 27.4$\pm 6.5$ & 50.0$\pm 16.0$ &   $3.7\pm 0.3$ & $0.899 (25 d.o.f.)$ \\
332 &power & $2.5\pm 0.1$  & - & - &   $3.0\pm 0.2$  & $5.711 (9 d.o.f.)$ \\
& bremss & $28.1\pm 1.2$  & - & - &  $2.9\pm 0.3$ & $1.216 (9 d.o.f.)$ \\
& pow*highe & $1.6\pm 0.4$  & 20.3$\pm 4.1$ & 30.9$\pm 9.8$ &   $2.9\pm 0.3$ & $0.932 (7 d.o.f.)$ \\
333 &power & $2.5\pm 0.1$  & - & - &   $1.8\pm 0.2$  & $2.697 (10 d.o.f.)$ \\
& bremss & $28.1\pm 1.9$  & - & - &  $1.7\pm 0.3$ & $1.764 (10 d.o.f.)$ \\
& bremss+power & 0.7$\pm 0.9$/20.0$\pm 4.5$ & - & -& $1.7\pm 0.3$& $0.932 (8 d.o.f)$ \\
335 &power & $2.6\pm 0.1$  & - & - &   $3.3\pm 0.3$  & $4.172 (10 d.o.f.)$ \\
& bremss & $30.4\pm 1.4$  & - & - &  $3.2\pm 0.3$ & $0.762 (10 d.o.f.)$ \\
& pow*highe & $1.7\pm 0.3$  & 21.7$\pm 4.9$ & 40.7$\pm 11.2$ &   $3.2\pm 0.3$ & $0.677 (8 d.o.f.)$ \\
385  &power & $2.7\pm 0.1$  & - & - &  $2.6\pm 0.3$  & $0.973 (27 d.o.f.)$ \\
& bremss & $26.2\pm 2.1$  & - & - &  $2.5\pm 0.3$ & $0.798 (27 d.o.f.)$ \\
& pow*highe & $1.4\pm 0.6$  & 13.6$\pm 15.5$ & 27.1$\pm 12.3$ &   $2.5\pm 0.3$ & $0.858 (25 d.o.f.)$ \\
386 & power & $2.6\pm 0.1$ & - & - &   $4.1\pm 0.3$ & $1.490 (29 d.o.f.)$ \\
& bremss & $28.3\pm 1.5$  & - & - &  $3.9\pm 0.3$ & $0.826 (29 d.o.f.)$ \\
& pow*highe & $1.6\pm 0.5$  & 19.0$\pm 5.1$ & 30.6$\pm 10.5$ &   $3.9\pm 0.3$ & $0.856 (27 d.o.f.)$ \\
389 & power & $2.8\pm 0.1$  &  - &-  & $2.4\pm 0.3$ &$1.052 (29 d.o.f)$ \\
 & bremss & 24.2$\pm 1.9$ & - & -& $2.2\pm 0.3$ & $1.059 (29 d.o.f)$ \\
 & bremss+power & 1.2$\pm 1.9$/18.6$\pm 5.5$ & - & -& $2.3\pm 0.3$& $0.911 (27 d.o.f)$ \\
395 & power & $2.6\pm 0.1$ & - & - &   $2.4\pm 0.3$ & $0.986 (29 d.o.f.)$ \\
& bremss & $30.4\pm 3.0$  & - & - &  $2.3\pm 0.3$ & $0.951 (29 d.o.f.)$ \\
& pow*highe & $1.9\pm 0.6$  & 18.0$\pm 11.2$ &50.7$\pm 31.8$ &   $2.3\pm 0.3$ & $0.977 (27 d.o.f.)$ \\
664 &power & $2.8\pm 0.1$  & - & - &   $2.4\pm 0.3$ &  $1.321 (10 d.o.f.)$ \\
& bremss & $22.1\pm 1.7$  & - & - &  $2.1\pm 0.3$ & $1.089 (10 d.o.f.)$ \\
& bremss+power & 0.9$\pm 1.3$/19.2$\pm 4.1$ & - & -& $2.2\pm 0.3$& $0.909 (8 d.o.f)$ \\
675 &power & $2.7\pm 0.1$  & - & - &   $2.6\pm 0.3$  & $1.580 (10 d.o.f.)$ \\
& bremss & $24.5\pm 1.6$  & - & - &  $2.4\pm 0.3$ & $0.922 (10 d.o.f.)$ \\
& pow*highe & $1.8\pm 0.4$  & 11.8$\pm 12.5$ & 35.9$\pm 15.8$ &   $2.5\pm 0.3$ & $0.916 (8 d.o.f.)$ \\
\hline
{\bf The quiescent states} &  &   & &  &   &  \\
\hline 145  & power & 2.7$\pm 0.6$ & - & -  & 0.47$\pm 0.11$ & $0.689 (5 d.o.f)$ \\
& bremss & $38.9\pm 25.8$  & - & - &  $0.43\pm 0.11$ & $0.577 (5 d.o.f.)$ \\
331 & power & $3.1\pm 0.3$ &- & - & $0.51\pm 0.09$ & $1.52 (4 d.o.f.)$ \\
& bremss & $13.6\pm 10.4$  & - & - &  $0.47\pm 0.09$ & $2.529 (4 d.o.f.)$ \\
384 & power & $2.6\pm 0.7$ & - & - & $0.19\pm 0.07$ & $1.21 (3 d.o.f.)$ \\
& bremss & $33.3\pm 11.5$  & - & - &  $0.19\pm 0.07$ & $1.02 (3 d.o.f.)$ \\
677 & power & $3.2\pm 0.6$ & - & - & $0.29\pm 0.09$ & $0.953 (4 d.o.f.)$ \\
& bremss & $15.3\pm 5.5$  & - & - &  $0.29\pm 0.09$ & $0.755 (4 d.o.f.)$ \\
\hline
\end{tabular}
\end{center}

\end{table*}

\end{document}